\documentclass[%
 aip,
 amsmath,amssymb,
 reprint,%
]{revtex4-1}

\usepackage{graphicx}
\usepackage{dcolumn}
\usepackage{bm}

\usepackage[utf8]{inputenc}
\usepackage[T1]{fontenc}
\usepackage{mathptmx}
\usepackage{xcolor}
\usepackage{float}

\begin{document}

\preprint{AIP/123-QED}

\title[]{Profilometry and stress analysis of suspended nanostructured thin films}

\author{Ali Akbar Darki}
\affiliation{Department of Physics and Astronomy, Aarhus University, DK-8000 Aarhus C, Denmark}
\author{Alexios Parthenopoulos}
\affiliation{Department of Physics and Astronomy, Aarhus University, DK-8000 Aarhus C, Denmark}
\author{Jens Vinge Nygaard}
\affiliation{Department of Engineering, Aarhus University, DK-8000 Aarhus C, Denmark}
\author{Aur\'{e}lien Dantan}
\email{dantan@phys.au.dk}
\affiliation{Department of Physics and Astronomy, Aarhus University, DK-8000 Aarhus C, Denmark}

\date{\today}

\begin{abstract}
The profile of suspended silicon nitride thin films patterned with one-dimensional subwavelength grating structures is investigated using Atomic Force Microscopy. We first show that the results of the profilometry can be used as input to Rigorous Coupled Wave Analysis simulations to predict the transmission spectrum of the gratings under illumination by monochromatic light at normal incidence and compare the results of the simulations with experiments. Secondly, we observe sharp vertical deflections of the films at the boundaries of the patterned area due to local modifications of the tensile stress during the patterning process. These deflections are experimentally observed for various grating structures and investigated on the basis of a simple analytical model as well as finite element method simulations.
\end{abstract}

\maketitle

\section{Introduction}

Suspended thin films are widely used in photonics and sensing applications. The use of subwavelength-structured films allows in particular for realizing ultracompact optical components with tailored properties and integrable in miniaturized devices~\cite{ChangHasnain2012,Zhou2014,Quaranta2018,Cheben2018}. For instance, dielectric or semiconductor films with subwavelength thickness and patterned with photonic crystal or subwavelength grating structures can be used as optical filters~\cite{Shuai2013,Wang2015}, couplers~\cite{Zhou2008,Penades2016,Kang2017}, reflectors~\cite{Bruckner2010}, lens~\cite{Fattal2010,Lu2010,Klemm2013}, polarizers~\cite{Mutlu2012}, spatial differentiators~\cite{Bykov2018,Dong2018,Yang2020,Parthenopoulos2020}, lasers~\cite{Boutami2007,Huang2008,Zhou2008,Wagner2016}, etc. On the other hand, exploiting the high mechanical quality of ultrathin suspended films also makes for highly sensitive opto- and/or electro-mechanical sensing devices~\cite{Midolo2018}.

A combination of lithography and etching is typically employed to pattern the films with such subwavelength structures and a precise knowledge of the topology and the post-fabrication stress of the structure is often desirable for understanding and optimizing the abovementioned applications. While the topology of patterned thin films deposited on a thick substrate can be straightforwardly realized by direct imaging of transverse cuts of the samples with e.g. a scanning electron microscope (SEM), the cutting of suspended patterned thin films is more delicate. A combination of metal coating followed by local illumination with a focused ion beam (FIB) and SEM imaging may be used to perform direct three-dimensional topology of freestanding films~\cite{Ierardi2014,Peltonen2016,Nair2019}, but it is in essence destructive. Atomic Force Microscopy (AFM), on the other hand, is a widely used method to characterize nanostructures and thus offers a natural noninvasive solution to the structural investigation of suspended nanostructured thin films.

In this article we report on the application of AFM to the profilometry of suspended, high tensile stress silicon nitride thin films patterned with one-dimensional subwavelength gratings (SWGs). The films of interest are 200-nm thick, 500 $\mu$m-square commercial Si$_3$N$_4$ membranes which initially possess a high tensile stress ($\sim$ GPa) as a result of the chemical vapor deposition process. We pattern these tensioned films with a subwavelength grating (period$\sim$ 800-850 nm) with various sizes and depths, using a combination of electron beam lithography (EBL) and plasma eching. The profile of the grating fingers and the overall deflection of the films are subsequently determined by means of appropriate AFM scans.

We first use the results of the profilometry of the grating fingers to predict the transmission spectrum of the SWGs under illumination by monochromatic light at normal incidence using Rigorous Coupled Wave Analysis (RCWA). The predicted spectra for different structures analyzed using this method show very good agreement with the experimentally measured ones.

The AFM profilometry also reveals sharp vertical deflections of the films at the borders of the patterned area due to the modification of the tensile stress during the patterning. These effects, well-known from e.g. high-contrast grating-based VCSEL~\cite{ChangHasnain2009} or NEMS~\cite{Gunay2012} fabrication, have interestingly also been observed with similar silicon nitride thin films exposed to a focused ion beam~\cite{Kim2006}. There, the FIB illumination resulted in local ion implantation, wich affected the tensile stress of the films and, thereby, its deflection. Here, the abrupt change in thickness results in a modification of the stress distribution and a subsequent deformation of the films. We thus first derive a simple one-dimensional model of a tensioned clamped plate with varying thickness to capture the essential features of the deflection. We then investigate experimentally the deflection of SWGs with different depths and sizes and compare the observations with the results of Finite Element Method (FEM) simulations. This provides us with useful insight into the residual stress distribution within the film, which is essential for the understanding of their mechanical properties, and ultimately may impose limitations on the nanostructuring of these freestanding films. Note that, due to the subwavelength nature of the structure and the highly local nature of the stress concentration, optical methods based on polarimetry~\cite{Ajovalasit2015,Capelle2017} would be difficult to apply here due to their limited optical resolution.

The methods and results of this work are generally relevant for the understanding of structural deformation of a wide range of nanostructured thin films. In particular, they may find interesting implications for the various suspended membrane resonators patterned with subwavelength grating or photonic crystal structures, which are currently widely used within optomechanics~\cite{Kemiktarak2012apl,Bui2012,Kemiktarak2012njp,Kemiktarak2014,Stambaugh2015,Yang2015,Norte2016,Bernard2016,Chen2017,Moura2018,Nair2019,Manjeshwar2020} and sensing~\cite{Guo2017,Naesby2018,Gartner2018,Cernotik2019,Dantan2020}.

\section{Suspended SWG fabrication and profilometry method}

\subsection{Fabrication}

\begin{figure}[h]
\centering
\includegraphics[width=\columnwidth]{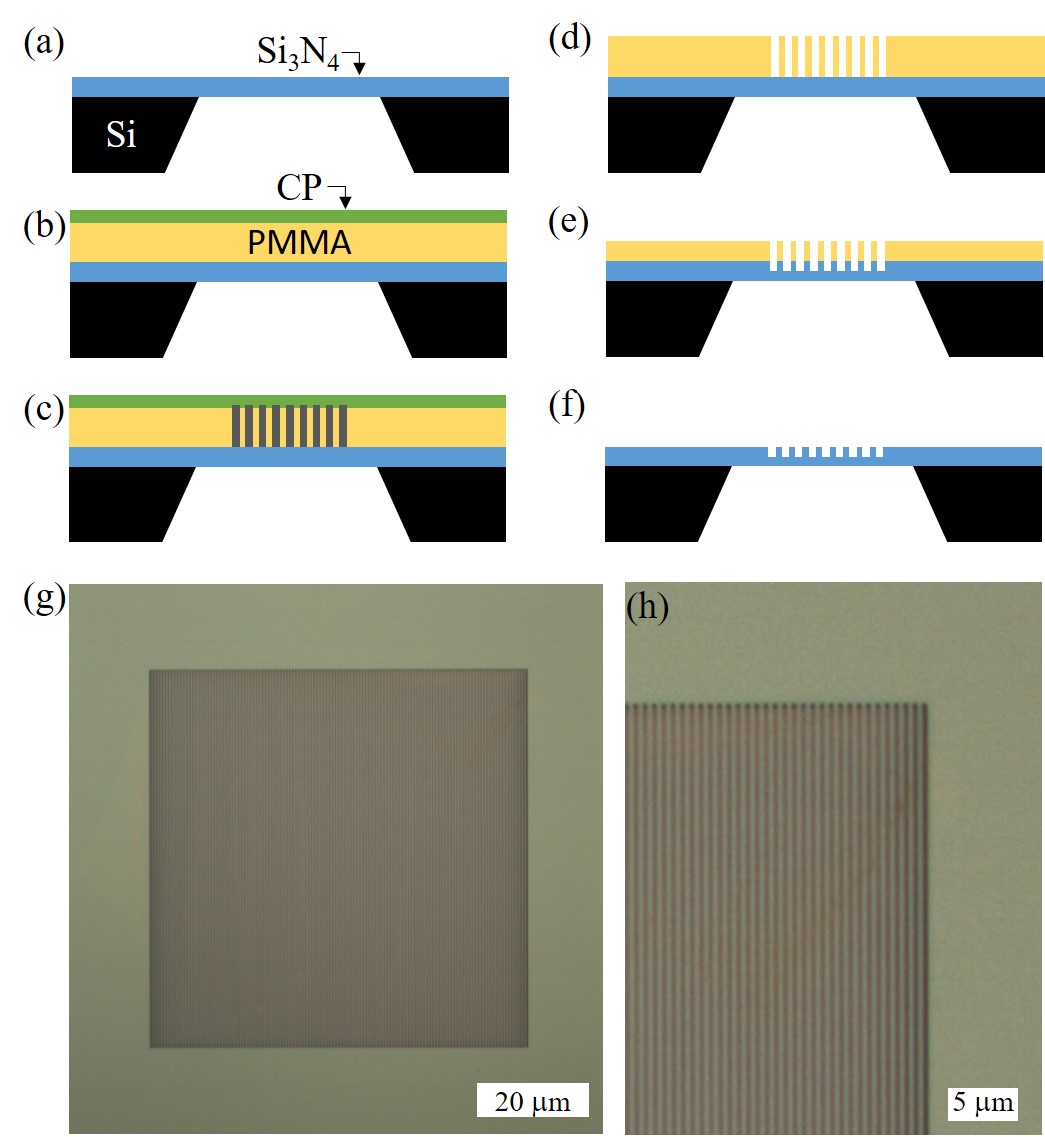}
\caption{Fabrication steps of a suspended SWG. (a) Commercial suspended Si$_3$N$_4$ membrane on Si. (b) Coating with PMMA and conductive polymer layer. (c) Electron Beam Lithography exposure. (d) Development. (e) Dry etching. (f) Suspended Si$_3$N$_4$ SWG. (g) Topview image of a 100 $\mu$m-square patterned structure. (h) Zoom on the top-right corner of the patterned structure of (g).}
\label{fig:fabrication}
\end{figure}

The nanostructured thin films are obtained following the recipe exposed in~\cite{Nair2019,Parthenopoulos2020}. In brief, we start with commercial (Norcada Inc., Canada), high tensile stress ($\sim$ GPa), stoichiometric silicon nitride film suspended on a silicon frame. The silicon nitride films used here are 200 nm-thick and deposited on a 5 mm-square, 200 $\mu$m-thick silicon frame. The lateral dimension of the suspended square membrane is 500 $\mu$m and the lateral size of the patterned structures varies between 50 $\mu$m and 400 $\mu$m. 

To fabricate the suspended SWG structure we follow the steps depicted in Fig.~\ref{fig:fabrication}(a)-(f). After oxygen plasma cleaning the samples are spin-coated with a PMMA layer and a conductive polymer layer. A square grating mask is written by EBL. The conductive polymer layer is subsequently removed by immersion in deionized water and the PMMA resist is developed in an water/IPA solution. The developed sample is then etched using reactive ion etching with C$4$F$_8$ and SF$_6$. The PMMA is removed in acetone and the sample is cleaned and dried with N$_2$. Figure~\ref{fig:fabrication}(g) shows a topview picture taken with an optical microscope of a patterned SWG. Such SWGs typically possess slented fingers, whose depth and wall slope may substantially vary depending the fabrication parameters and the etching time used. For reasons which will be discussed further, we typically etch between 20\% and 75\% of the thickness of the membrane in these experiments. The resulting SWG is thus a one-dimensional grating with trapezoidal fingers and an underlying silicon nitride layer, whose thickness is somehow comparable to the finger height.

Images such as those shown in Figs.~\ref{fig:fabrication}(g) and (h) only provide partial information on the resulting SWG dimensions--the grating period and, to some extent, the mean finger width, but do not allow a precise determination of the finger depth or the shape of the wall profiles. Since one deals with suspended films during the whole process, cutting of the structure is delicate and, while it can be performed using e.g. a Focused Ion Beam, it is destructive. As a noninvasive alternative to the FIB cutting method used in~\cite{Nair2019} to characterize this type of SWGs, we investigate Atomic Force Microscopy in the following.

\subsection{AFM profilometry method}
\label{sec:AFMmethod}

\begin{figure}[h]
\centering
\includegraphics[width=\columnwidth]{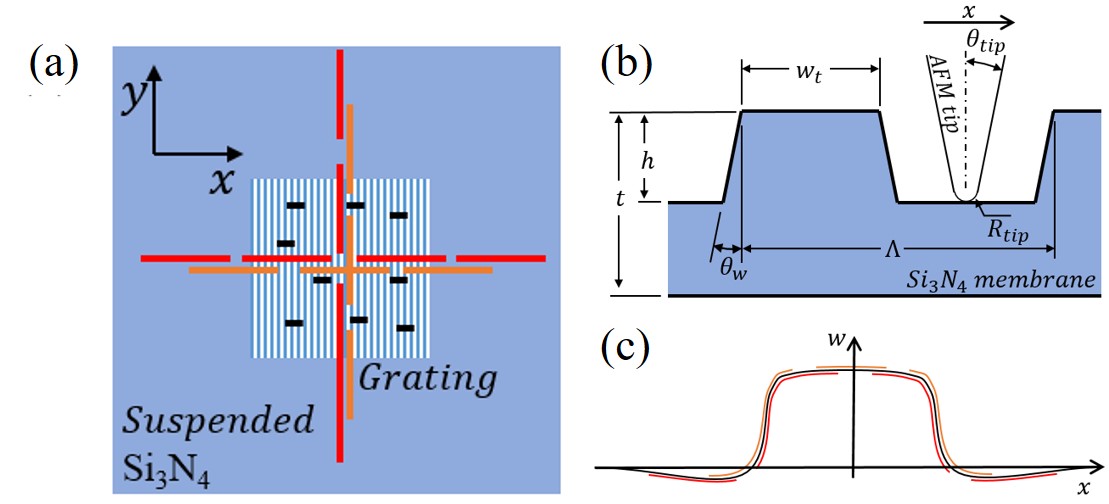}
\caption{(a) Scanning methods: the short, black and horizontal lines represent short (10 $\mu$m) scans in the $x$-direction, while the long, red/orange, horizontal/vertical lines represent longer scans (100 $\mu$m) in either the $x$- or the $y$-directions. (b) Short scans: trapezoidal SWG profilometry. (c) Long scans: overall patterned area profilometry to evaluate the membrane vertical deflection $w$, e.g. in the $x$-direction.}
\label{fig:scanning}
\end{figure}

The instrument used in this work to measure the suspended film profile is a Brucker Dimension Edge AFM. We made use of Brucker RTESPA-300 AFM tips, which are pyramid-shaped with a specified tip radius $R_{tip}=(10\pm2)$ nm and side angles $\theta_{tip}=(17.5\pm2)^{\circ}$, as shown in Fig.~\ref{fig:scanning}(b). Two types of scans in tapping mode were performed in this study: {\it (i)} {\it short} scans (10 $\mu$m) perpendicular to the grating fingers in different regions of the patterned area, as depicted in Fig.~\ref{fig:scanning}(a), and {\it (ii)} {\it long} scans (100 $\mu$m) spanning both the patterned and unpatterned areas and in the directions either parallel or perpendicular to the grating fingers (Fig.~\ref{fig:scanning}(c)).

The goals of the short scans are both to characterize the profile of the grating fingers in order to precisely extract the relevant SWG geometrical parameters (period $\Lambda$, top finger width $w_t$, finger height $h$ and, insofar as it is larger than the AFM tip angle $\theta_{tip}$, the wall slope angle $\theta_w$) and to assess the homogeneity of the structure. These scans are thus performed at the lowest possible speed (2 $\mu$m/s). To improve the accuracy of the measurements the error due to the tip-shape~\cite{Hubner2003} is compensated after fitting the measured profile with a trapezoidal profile.

The goal of the long scans is to assess the overall vertical deflection of the membrane (Fig.~\ref{fig:scanning}(c)), resulting from the SWG patterning which critically affects the residual tensile stress of the film. The individual scans are typically not long enough to cover the width of the membrane or of the patterned area, but a careful overlap between consecutive scans allows for binding the resulting curves together to capture the overall deformation. Since the scans are long, the circular arc error~\cite{Lianqing2008} is not negligible (19 nm for a 100 $\mu$m scan) and is compensated before binding the curves.

\section{Profilometric and optical characterization of the SWG}

\begin{figure}[h]
\centering
\includegraphics[width=\columnwidth]{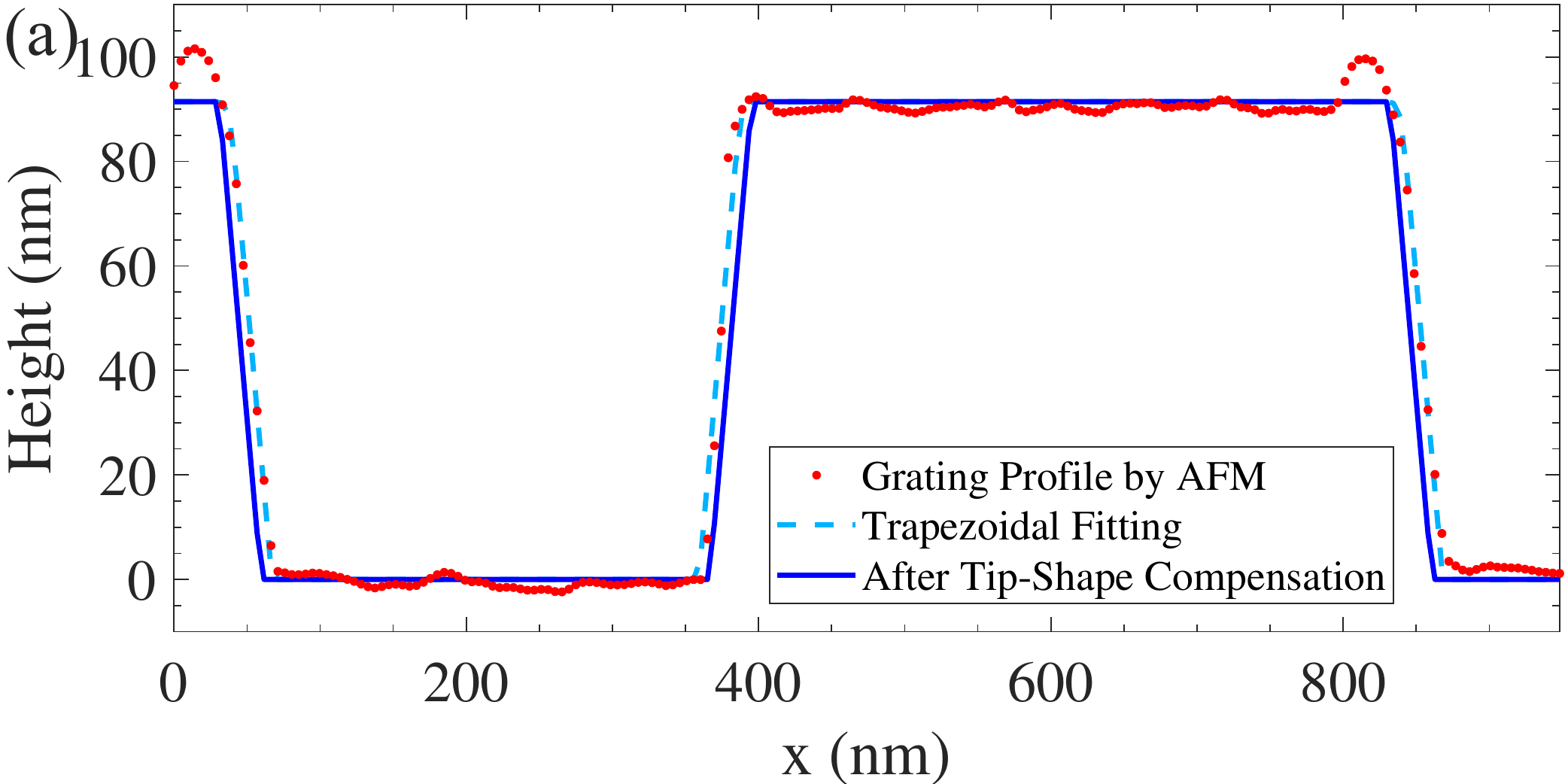}\\
\includegraphics[width=\columnwidth]{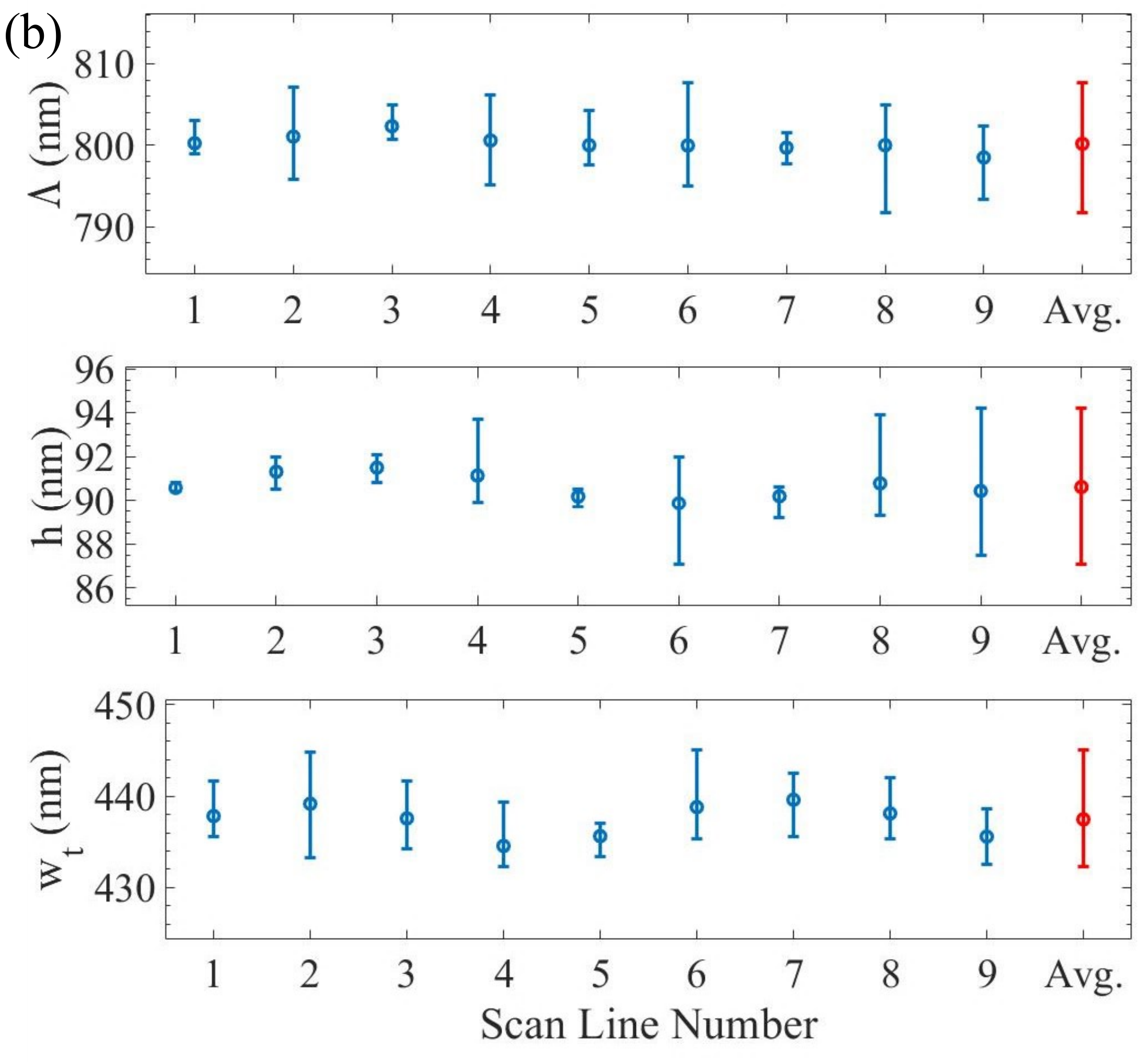}
\caption{(a) Results of a short scan performed on sample B2 (Fig.~\ref{fig:fabrication}(g)). Red dots: experimental data. Dashed cyan line: result of a fit to a trapezoidal profile. Plain blue line: profile extracted after AFM tip curvature correction. (b) Profilometry results for 9 short scans in different parts of sample B2.}
\label{fig:grating_profile}
\end{figure}

An example of profilometry performed using short scans in 9 different positions of the SWG shown in Fig.~\ref{fig:fabrication}(g) (sample B2) is given in Fig.~\ref{fig:grating_profile}(a), where the red dots show the experimentally measured profile, the plain blue line the result of a fit to a trapezoidal profile, such as defined in Fig.~\ref{fig:scanning}(b), and the dashed blue line the profile extracted from the trapezoidal fit taking into account the AFM tip curvature. Such an analysis allows for locally extracting the grating period $\Lambda$, the top finger width $w_t$, the finger height $h$ and the grating wall slope angle $\theta_w$, insofar as it is larger than that of the AFM tip $\theta_{tip}$. The mean finger width is then $w_m=w_t+h\tan\theta_w$. The results of the parameters extracted in 9 different parts of the grating are reported in Fig.~\ref{fig:grating_profile}(b) and show good homogeneity of the measured profiles within the uncertainties of the individual local measurements.

\begin{figure}[h]
\centering
\includegraphics[width=\columnwidth]{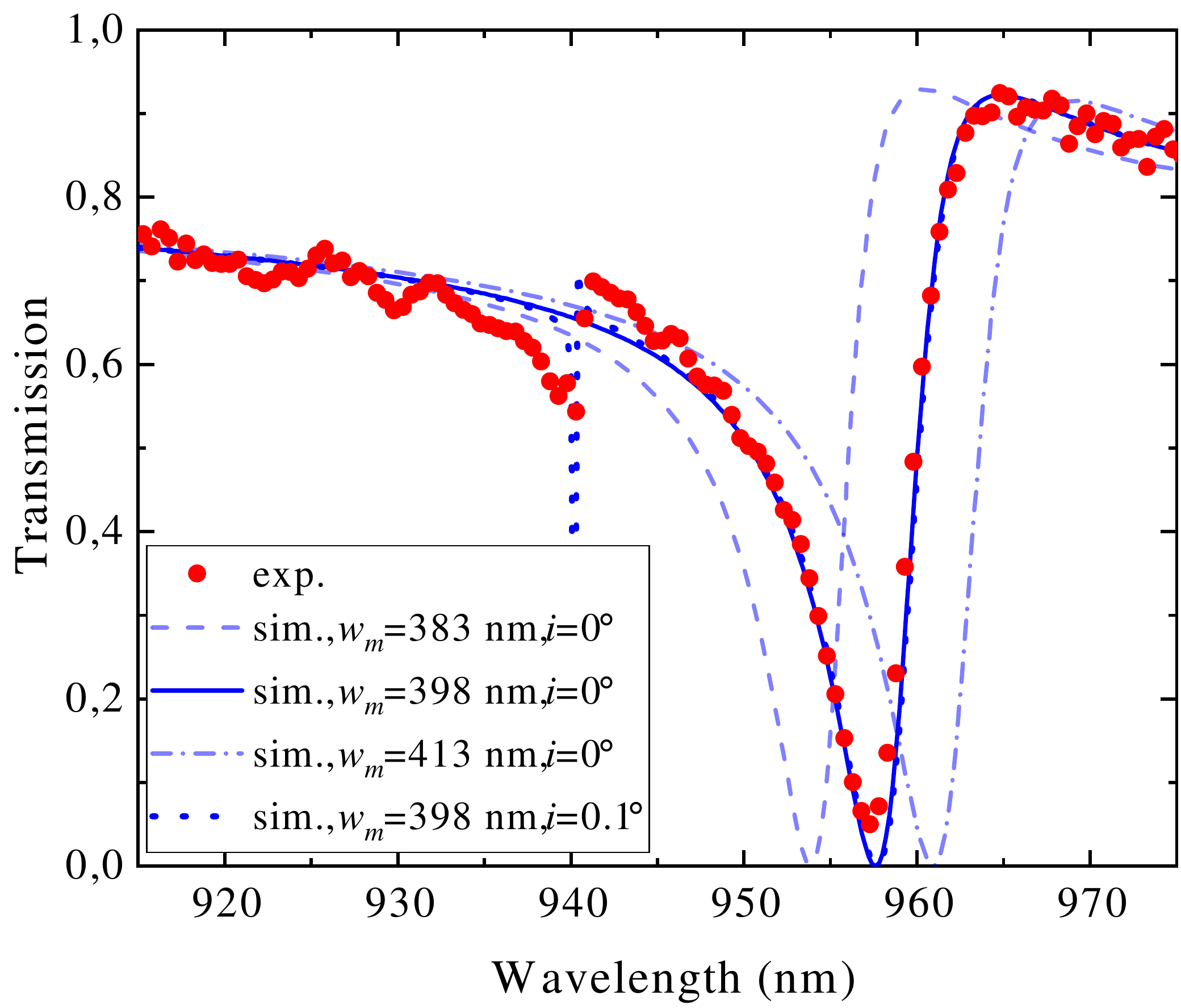}
\caption{Measured normalized transmission spectrum (red dots) of sample A2 (see text and Table \ref{tab:parameters} for parameters) illuminated at normal incidence with TM-polarized light and exhibiting a high-reflection Fano resonance around 957.5 nm. Plain blue line: predictions of RCWA simulations using the average grating parameters determined by the AFM scans and $w_m=398$ nm ($\theta_w\simeq \theta_{tip}/2$). Dashed and dot-dashed blue lines: RCWA simulation results for $w_m=383$ nm ($\theta_w=0$) and $w_m=413$ nm ($\theta_w=\theta_{tip}$). Dotted blue line: RCWA simulations for $w_m=398$ nm for a small angle ($i=0.1^{\circ}$) with respect to normal incidence of the incident light to evidence the presence of a second Fano resonance around 939 nm.}
\label{fig:transmission}
\end{figure}

The determined geometrical grating parameters can be used as input to RCWA simulations in order to predict the transmission of the grating under monochromatic illumination with polarized light. The refractive index (2.00) and thickness ($t=203$ nm) of the unpatterned films are independently determined using ellipsometry and used as well as input parameters for the simulations. The RCWA simulations are performed using the MIST software (Modeled Integrated Scattered Tool~\cite{MIST}), by discretizing the infinite one-dimensional structure in 20 layers and using a 25 mode basis. The simulations assume incident monochromatic plane wave illumination with linearly polarized light. 

In Fig.~\ref{fig:transmission}, the predicted normalized transmission spectrum for TM-polarized light impinging at normal incidence on another SWG (sample A2), with a 200 $\mu$m-square patterned area and with parameters slightly different than that of Fig.~\ref{fig:grating_profile}, $\Lambda=(857\pm 6)$~nm, $h=(103\pm 1)$~nm and $w_t=(383\pm 7)$ nm, are compared with the experimentally measured one. The experimental spectrum shows a Fano resonance around 957.5 nm, corresponding to a high reflectivity resonance. Such a Fano resonance results from the interference between a resonant guided mode in the structure and incoming light with a specific wavelength and polarization~\cite{Fan2002}, and the SWG parameters were chosen so that this resonance should be observable within the available laser wavelength tuning range.

The experimental spectrum is obtained by gently focusing the spatially filtered light from a tunable external cavity diode laser (Toptica DLC Pro) and measuring the transmitted light power with a photodetector referenced to the incident power, as detailed in~\cite{Nair2019,Parthenopoulos2020}. The spectrum in presence of the SWG in the range 915-975 nm is then normalized by that without it. The large size of the SWG structure--$(200\;\mu$m)$^2$ for sample A2--allows for operating with a relatively large beam waist (spotsize$\sim$ 130 $\mu$m), thus minimizing collimation effects due to the fact that the incident light beam is not a plane wave, but a Gaussian beam, i.e. a superposition of plane waves with different incidence angles. We verified that, with this focusing, collimation effects negligibly affect the position and width of the observed Fano resonance.

Figure~\ref{fig:transmission} shows as well the results of the RCWA simulations using as input the mean period, top finger width and finger heigth determined by the short AFM scans and for different wall slope angles, $\theta_w=0$, $\theta=\theta_{tip}$ and $\theta_w\simeq\theta_{tip}/2$. The excellent agreement between the experimental and simulated spectra--both in terms of the position of the Fano resonance and its width--obtained for a mean finger width of $w_m=398$ nm suggests that the walls are thus tilted with an angle which is approximately half of the specificed AFM tip angle in the case of this sample. Similar behavior was observed with the other samples used in this work, although the wall slope angle may vary depending on the fabrication process (see table~\ref{tab:parameters} and Appendix). 

It is also interesting to note that a second, smaller Fano resonance is also visible around 940 nm. This resonance arises from the coupling of the incident light to a grating guided mode possessing a different symmetry than that of the guided mode resonantly interfering with the incident light at 957.5 nm. For a perfectly homogeneous and infinite SWG and for a suitably polarized plane wave at normal incidence, the symmetry of the guided modes only allows scattering into a guided mode with a certain parity and, therefore, the observation of a single resonance only, as shown by the simulated spectra with $i=0^\circ$ in Fig.~\ref{fig:transmission}. At oblique incidence, however, excitation of a second guided mode with different parity becomes possible, leading to the apparition of a second Fano resonance (dotted blue curve)~\cite{Fan2002,Bykov2015}. We refer the reader to~\cite{Parthenopoulos2020} for a detailed study of oblique incidence effects with this kind of SWGs. Of interest here is that the nonideality of the grating and/or the Gaussian nature of the incident beam typically makes this second resonance visible in the experimental spectrum, even at normal incidence. The correct prediction of the position of this second resonance based on the parameters extracted from the AFM profilometry thus also supports the accuracy of the method.

We performed such scans and analysis for grating structures with various parameters, in particular with different patterned area sizes and depths. When the Fano resonances were in the available wavelength range of the laser used for the transmission measurements, the finger wall angle, and thereby the mean finger width, were determined as discussed previously. Additional examples of measured and simulated transmission spectra are given in the Appendix. The resulting parameters for the SWGs used in this work are reported in Table~\ref{tab:parameters}.

\begin{table}[h]
\caption{\label{tab:parameters}Geometrical parameters of the SWGs used in this work.}
\begin{ruledtabular}
\begin{tabular}{cccccc}
Sample & b & $\Lambda$ & $h$ & $w_t$ & $w_m$\\
A1 & 200 & 856 & 40 & 409 & 421\\
A2 & 200 & 857 & 103 & 383 & 393 \\
A3 & 200 & 858 & 153 & 394 &  411\\
B1 & 50   & 820 & 107 & 487 & 487\\
B2 & 100 & 800 & 91 & 439 & 453\\
B3 & 200 & 801 & 90 & 416 &  434\\
\end{tabular}
\end{ruledtabular}
\end{table}


\section{Deflection and stress investigations}

We now turn to the investigation of the overall deflection of the patterned membrane using the long AFM scans introduced in Sec.~\ref{sec:AFMmethod}, and, in particular, to what occurs at the borders of the etched area. We start by discussing such deflection effects within the frame of a simple one-dimensional model of a thin, tensioned membrane with varying thickness. This has the advantage of allowing for transparent analytical predictions capturing the essential of the physics, before discussing the more complex two-dimensional SWG case on the basis of experimental measurements and finite element simulations.

\subsection{One-dimensional model}

\begin{figure}[h]
\centering
\includegraphics[width=0.75\columnwidth]{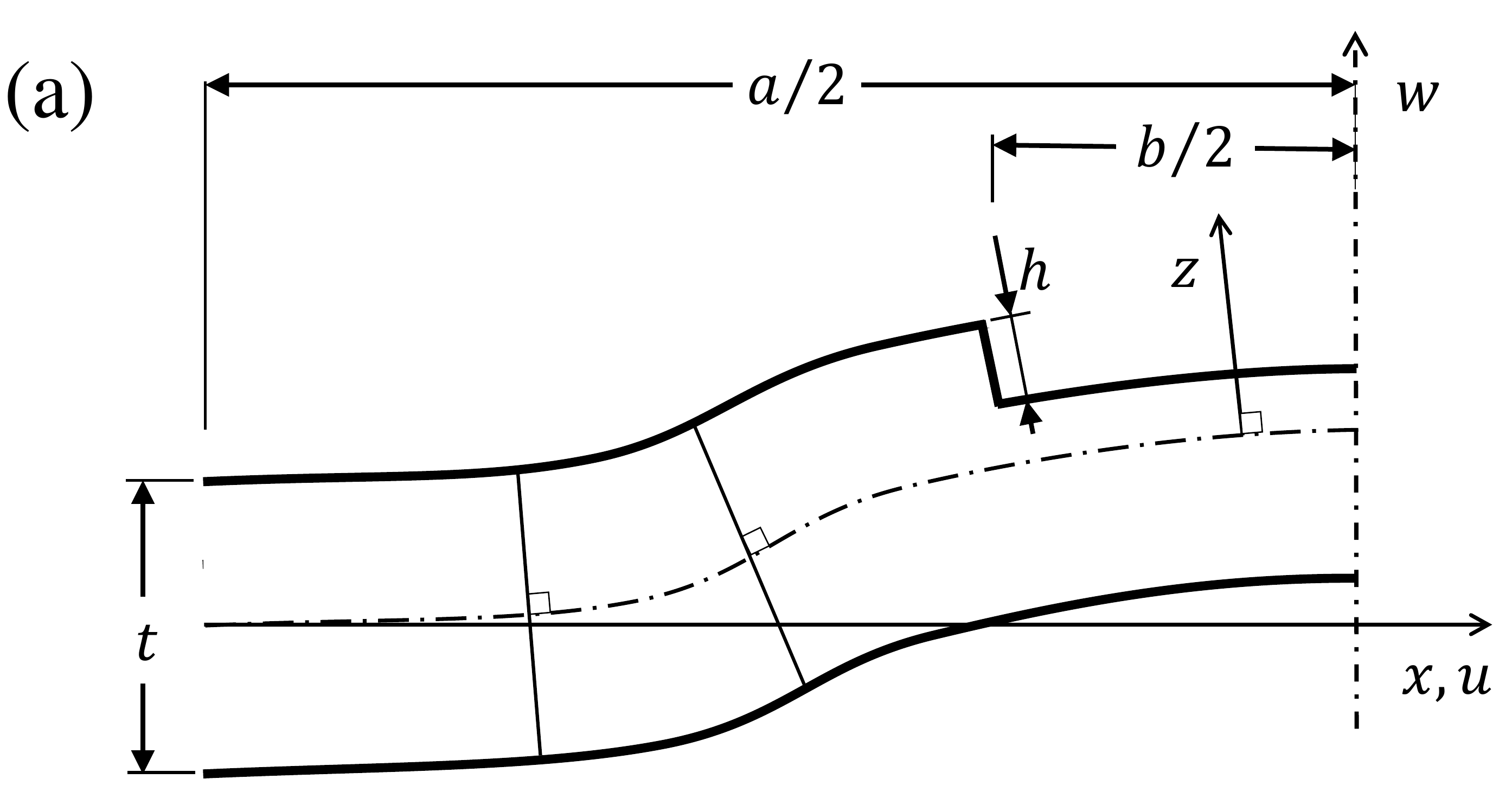}
\includegraphics[width=0.75\columnwidth]{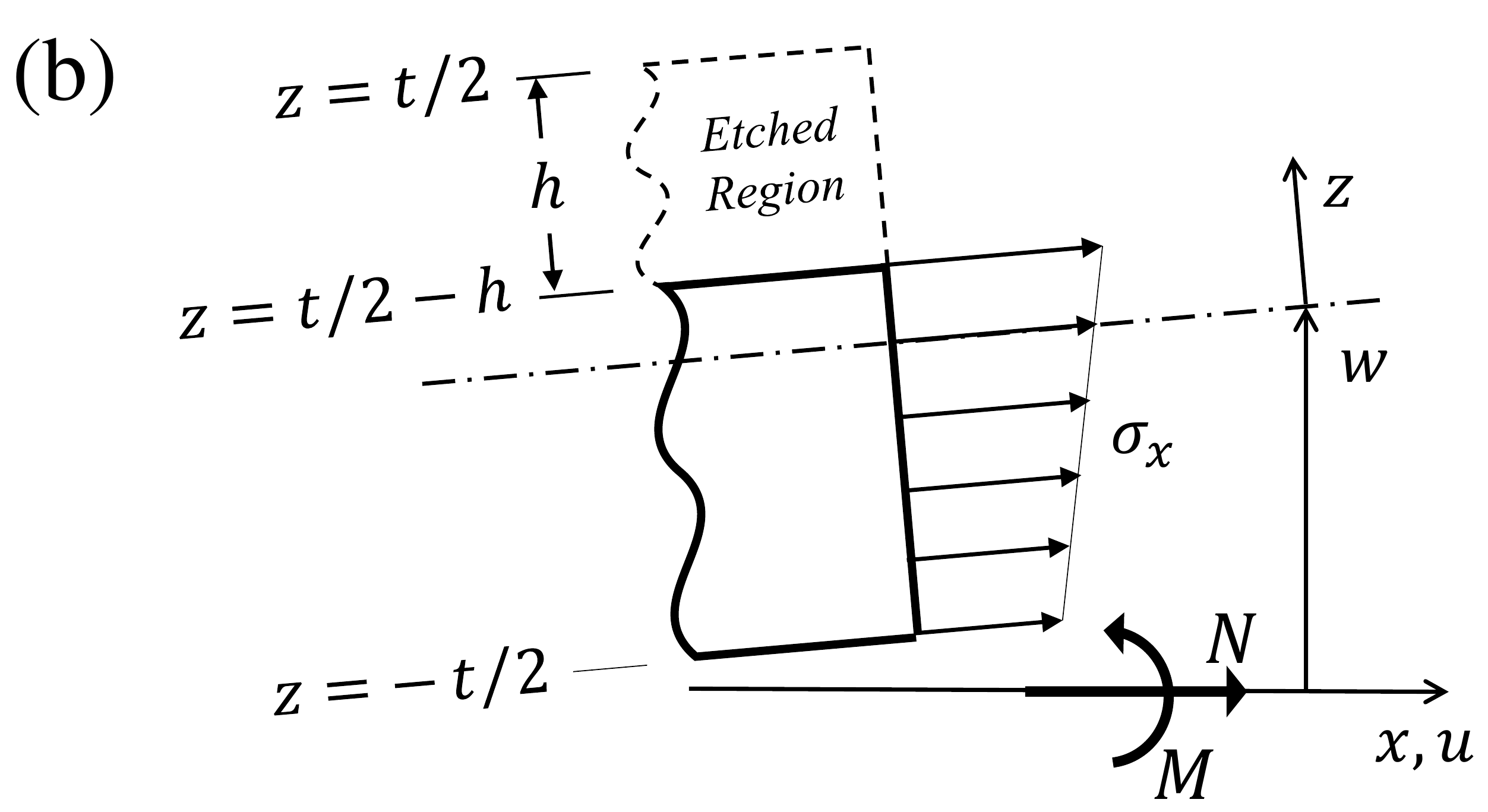}
\caption{(a) Schematic membrane deflection in the 1D model. The profile is symmetric with respect to the $y$-axis. (b) Conventions for the tensile stress and line force and moments used in the 1D model.}
\label{fig:1D_principle}
\end{figure}

We start by considering a clamped plate, of length $a$ in the $x$-direction and infinite in the $y$-direction. The plate is clamped at $x=-a/2$ and $x=a/2$ and has initially a uniform residual tensile stress $\sigma_0$. As depicted in Fig.~\ref{fig:1D_principle}, a height $h$ is then removed from the central part with length $b$ of the plate, so that the plate thickness in the region $-b/2<x<b/2$ becomes $t-h$, whereas it remains $t$ in the regions $-a/2<x<-b/2$ and $b/2<x<a/2$.  Due to the symmetry with respect to the center of the plate at $x=0$, we focus on the $-a/2<x<0$ region. We denote the plate's elasticity (Young's) modulus and Poisson ratio by $E$ and $\nu$, respectively. Denoting by $w_i$ the equilibrium deflections of the membrane in the $z$-direction and $u_i$ the displacement fields in the non-etched ($i=1$) and etched ($i=2$) regions, the equilibrium relations for the line forces, $N_i$, and line moments, $M_i$, read~\cite{Timoshenko1959}
\begin{align}
N_1&=\tilde{E} u'_1 t,\label{eq:N1}\\
N_2&=\tilde{E}\left(u'_2+w_2''\frac{h}{2}\right)(t-h),\\
M_1&=D_1w_1''-N_1w_1,\\
M_2&=D_2w_2''-N_2\left(w_2-\frac{h}{2}\right),\label{eq:M2}
\end{align}
where 
\begin{equation}
\tilde{E}=\frac{E}{1-\nu^2},\hspace{0.2cm}D_1=\frac{\tilde{E}t^3}{12},\hspace{0.2cm}D_2=\frac{\tilde{E}(t-h)^3}{12},
\end{equation} 
and the prime and double prime denote the first- and second-derivative with respect to $x$, respectively.

We set the line forces and line moments to be equal at equilibrium:
\begin{equation}
N_1=N_2\equiv N_0\hspace{0.2cm}\textrm{and}\hspace{0.2cm} M_1=M_2\equiv M_0,
\end{equation}
and assume the boundary conditions
\begin{align}
w_1(-a/2)&=0,\hspace{0.2cm}w_1'(-a/2)=0,\hspace{0.2cm}w_2'(0)=0,\\
u_1(-a/2)&=-\sigma_0a/(2\tilde{E}),\hspace{0.2cm}u_2(0)=0,
\end{align}
as well as continuity of the displacements and their derivatives at the border $x=-b/2$
\begin{align}
w_1(-b/2)&=w_2(-b/2),\\
w_1'(-b/2)&=w_2'(-b/2),\\
u_1(-b/2)&=u_2(-b/2).
\end{align}
This allows us to solve (\ref{eq:N1}-\ref{eq:M2}) analytically and we obtain
\begin{align}
\label{eq:w1} w_1(x)&=\frac{h}{2}\frac{1}{C_1+\alpha C_2}\left[\cosh(K_1(x+a/2))-1\right],\\
\label{eq:w2} w_2(x)&=\frac{h}{2}\left[1-\frac{1}{C_1+\alpha C_2}-\frac{\alpha}{C_1+\alpha C_2}\cosh(K_2x)\right],
\end{align}
with
\begin{align}
K_i&=\sqrt{\frac{N_0}{D_i}},\hspace{0.2cm}S_i=\sinh\beta_i,\hspace{0.2cm}C_i=\cosh\beta_i\hspace{0.1cm}(i=1,2),\\
\alpha &=\frac{K_1 S_1}{K_2S_2},\hspace{0.2cm}\beta_1=K_1(a-b)/2,\hspace{0.2cm}\beta_2=K_2b/2.
\end{align}
$N_0$ can be obtained as an implicit solution of a complex equation, but, in the thin plate and large $\beta_i$'s limits relevant here, it can be shown to be approximately given by
\begin{equation}
N_0\simeq\sigma_0 t\frac{a(t-h)}{a(t-h)+bh}.
\end{equation}
The $x$-component of the local stress in the different regions can then be computed from
\begin{equation}
\sigma_{x,i}=\tilde{E}\left(u_i'(x)-z w_i''(x)\right),
\end{equation}
where the $w_i$'s are given by Eqs.~(\ref{eq:w1}-\ref{eq:w2}) and the $u_i'$'s by
\begin{align}
u_1'(x)&=\frac{N_0}{\tilde{E}t},\\
u_2'(x)&=\frac{N_0}{\tilde{E}(t-h)}+\frac{h^2}{4}\frac{\alpha K_2^2}{C_1+\alpha C_2}\cosh (K_2x).
\end{align}

\begin{figure}[h]
\centering
\includegraphics[width=\columnwidth]{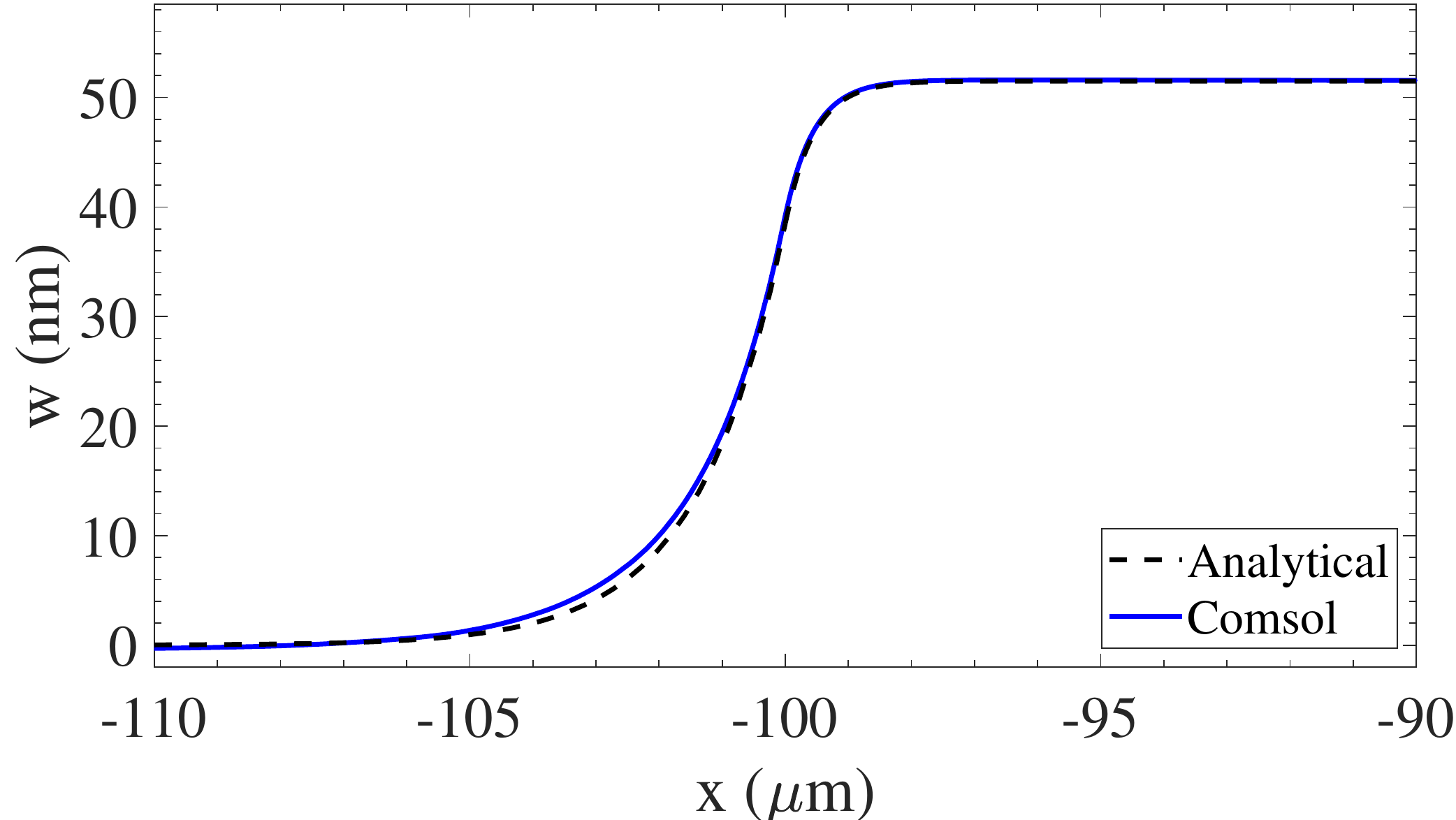}\\
\includegraphics[width=\columnwidth]{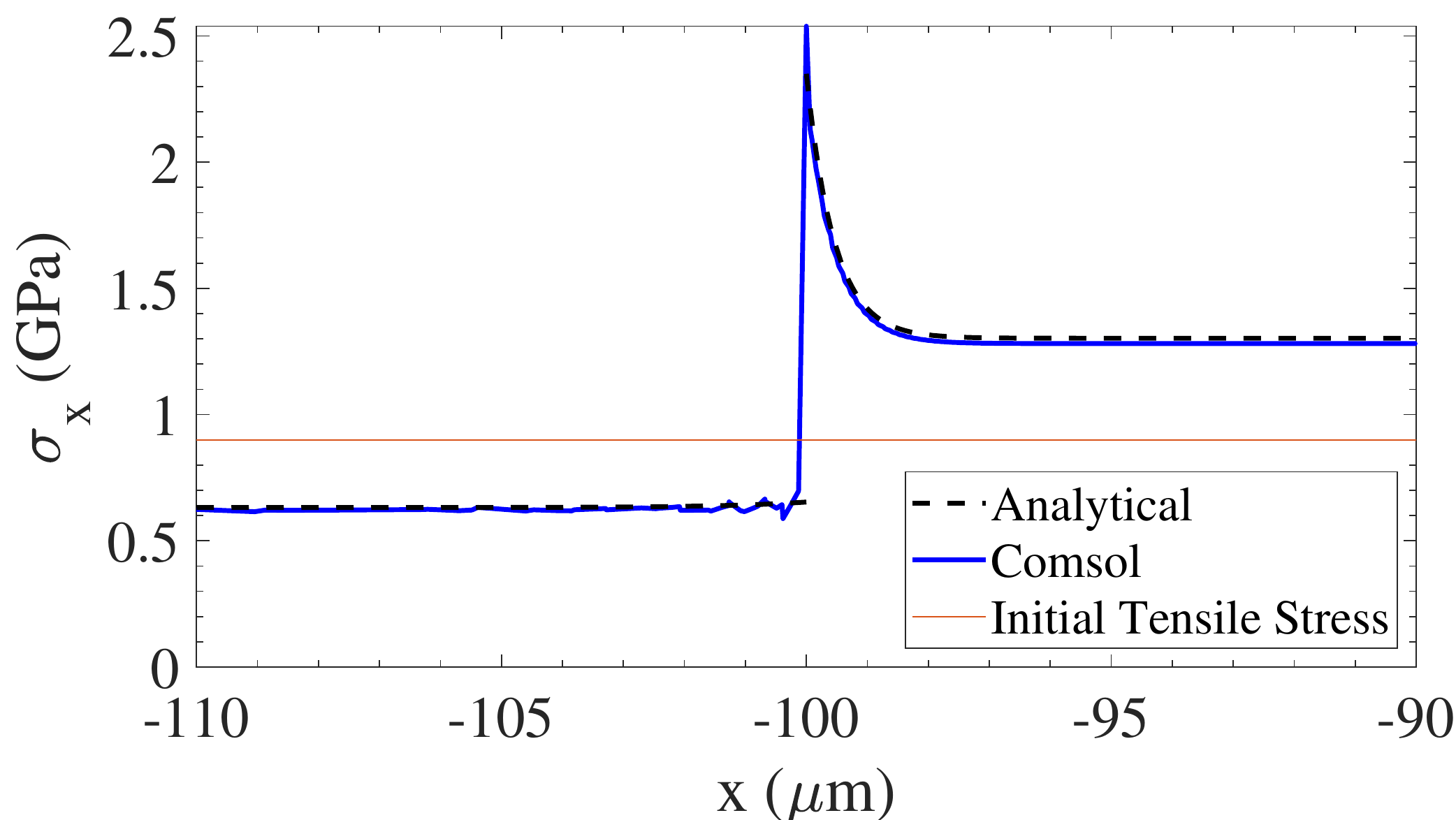}
\caption{Variations of $w(x)$ (top) and $\sigma_x(x,z=t/2-h)$ (bottom), as a function of $x$, for an infinitely long membrane in the $y$-direction with $E=316$ GPa, $\nu=0.298$, $\sigma_0=0.9$ GPa and dimensions $a=500$ $\mu$m, $b=200$ $\mu$m, $t=200$ nm, $h=103$ nm. Dashed: analytical 1D model predictions. Plain: 3D FEM simulation results for a membrane with the same parameters, but with length 500 $\mu$m in the $y$-direction.}
\label{fig:1D_results}
\end{figure}
 
Figure~\ref{fig:1D_results}(a) shows the predicted variations along $x$ of the membrane deflection in the $z$ direction for a "one-dimensional" membrane (infinite in the $y$-direction) with $E=316$ GPa, $\nu=0.298$, $\sigma_0=0.9$ GPa and dimensions $a=500$ $\mu$m, $b=200$ $\mu$m, $t=200$ nm, $h=103$ nm. A sharp upward deflection is observed at the border of the patterned area. This deflection occurs on length scales on each side of the border given by $1/K_i$ and leads to a height change equal to $h/2$ in the limit of large $\beta_i$'s. Figure~\ref{fig:1D_results}(b) shows the corresponding variations of the $x$-component of the stress evaluated at $z=t/2-h$, i.e. at the top of the etched area. The stress increases in the etched area due to the reduction in thickness of the membrane. A sharp decrease or increase in the stress can be also observed on each side of the border of the etched area. It can be shown that the stress always increases and is maximal on the etched side and strongly depends on the etching depth. The predictions of the analytical model were verified by carrying out full three-dimensional finite element simulations using COMSOL, as will be detailed in the next section. The results for a membrane with a length 500 $\mu$m in the $y$-direction are given in Fig.~\ref{fig:1D_results} and show excellent agreement with the predictions of the one-dimensional model. Let us also note, though, that the maximal value in tensile stress observed at the edge of the patterned area in the simulations is quite dependent on the curvature of the corner at $z=t/2-h$. This is a consequence of the stress concentration in this region, phenomenon which is captured by the FEM simulations, but not by the linear stress profile assumed in the analytical model (Fig.~\ref{fig:1D_principle}(b)). Fig.~\ref{fig:1D_results} shows that, assuming a radius of curvature of 20 nm in the simulations, stress concentration is negligible and good agreement is obtained between the analytical model predictions and the simulations.

\subsection{Two-dimensional square SWGs: simulations and experimental results}

Two important elements are missing in the previous simplified one-dimensional analysis in order to compare with the experimental observations of the fabricated SWGs. First, the patterned areas are square, so that the stress release is not unidimensional when etching. Second, the grating fingers introduce a directional asymmetry, so one expects different deflections in the directions perpendicular or parallel with the grating fingers. While a generalization of the one-dimensional analytical model of the previous section may be possible, we leave it for future investigations and investigate below the case of two-dimensional square SWGs on the basis of experimental AFM measurements and of full three-dimensional finite element simulations.

\subsubsection{Sample A2}

\begin{figure}[h]
\centering
\includegraphics[width=\columnwidth]{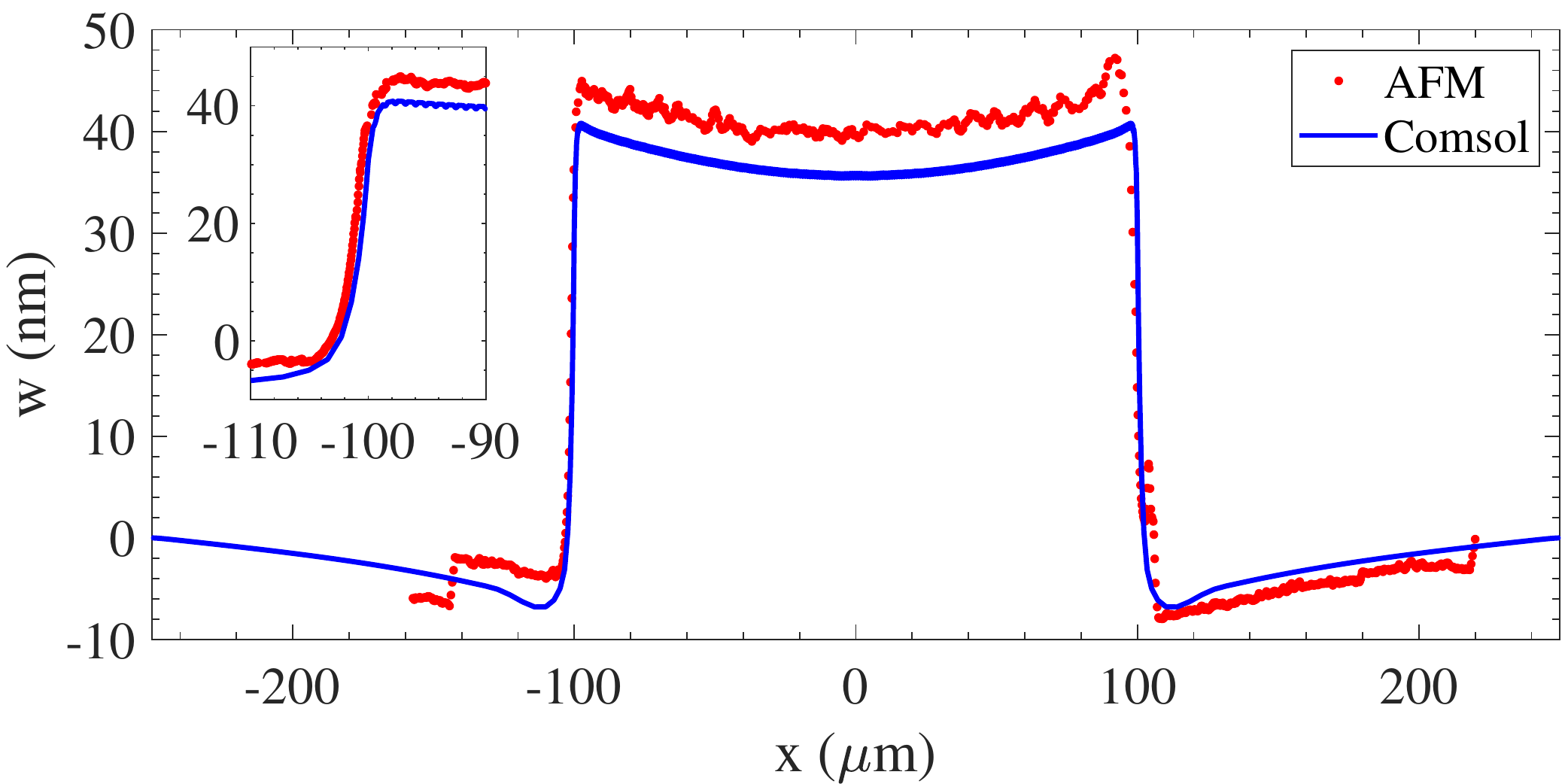}
\includegraphics[width=\columnwidth]{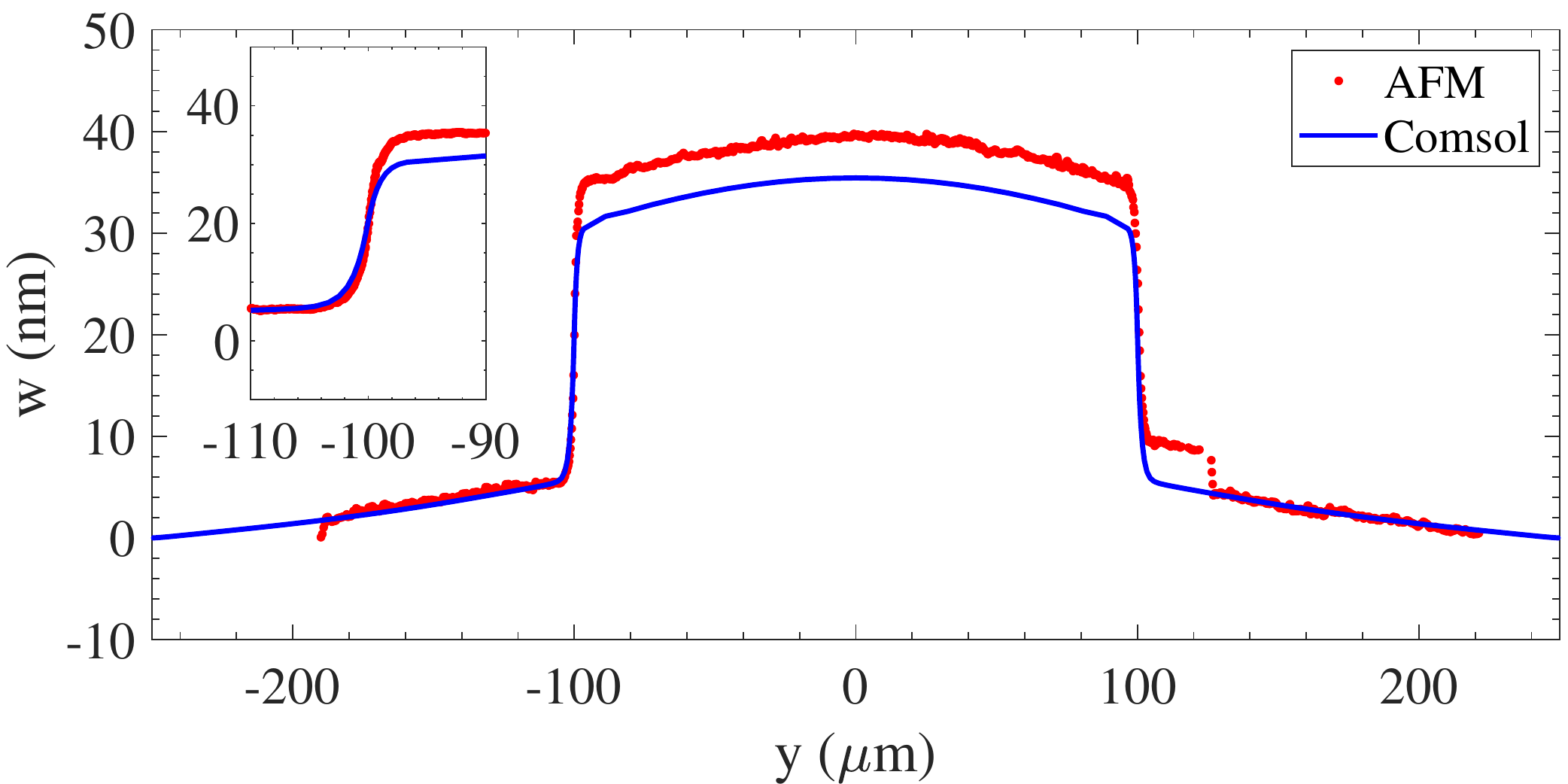}
\caption{Vertical deflection measured in the directions perpendicular (top) and parallel (bottom) to the grating fingers for sample A2. The insets show an enlarged view of the deflection region at the border of the patterned area.}
\label{fig:bucklingG07}
\end{figure}

Figure~\ref{fig:bucklingG07} shows the deflection measured in the course of long scans in the directions perpendicular ($x$) and parallel ($y$) to the grating fingers for the sample whose transmission spectrum is shown in Fig.~\ref{fig:transmission}. The two scans cross at the center of the square grating. For readability reasons, only the slowly varying envelope of the deflection--and not the individual grooves--are shown for the scan in the $x$-direction. Clear upward deflection is observed at the border of the patterned area in both directions. A stronger and sharper deflection is seen in the $x$-direction than in the $y$-direction, though. This can be expected, since the change in thickness is more abrupt at the border in the $x$-direction, and the stress modification is therefore stronger at the edge parallel with the fingers. Consequently, while the difference in height between the bottom of the depression before the edge of the first groove and its maximum elevation is about $h/2$ in the $x$-direction, the height change is less in the $y$-direction due to the effectively lower effective etching depth experienced in this $y$-direction. The deflection in the $y$-direction thus increases monotonically across the border to reach its maximum at the center of the patterned area. In contrast, in the $x$-direction, the deflection slowly decreases from the edge of the patterned area towards the center of the structure to match the maximal deflection in the $y$-direction there. The difference in height change in both directions also explains why a slight downward deflection is observed in the $x$-direction as one approaches the edge of the patterned area. 

These observations are well-corroborated by FEM simulations which were carried out using a full three-dimensional meshing of the clamped and pretensioned structure in COMSOL~\cite{Zienkiewicz2014}. The results of the FEM simulations using the geometrical grating parameters determined with short AFM scans and assuming an initial tensile stess of 0.9 GPa are also shown in Fig.~\ref{fig:bucklingG07}. Overall, the simulation results agree well with the experimental measurements. While the total deflection in the etched area is slightly underestimated by the simulations, the rapid deflections at the edge and the overall curvatures observed in the experiments are well-reproduced by the simulations, clearly showing how the deflection develops in the different directions. 

\begin{figure}[h]
\centering
\includegraphics[width=\columnwidth]{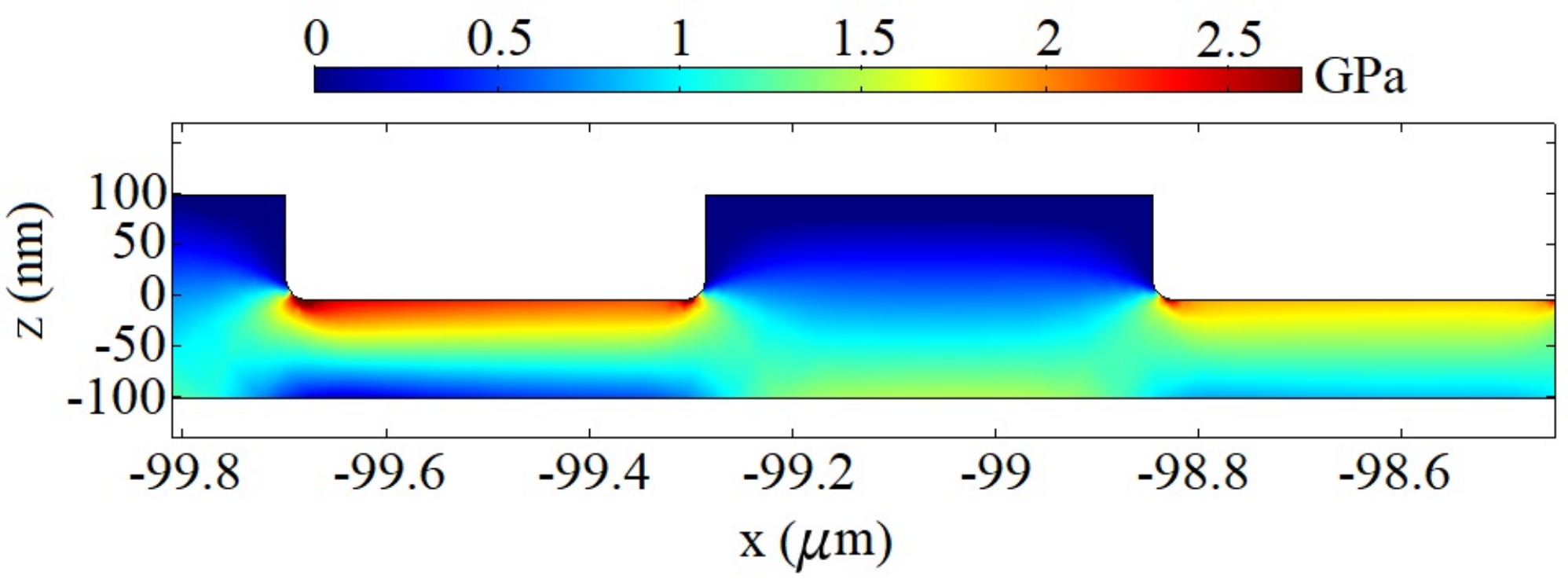}
\includegraphics[width=\columnwidth]{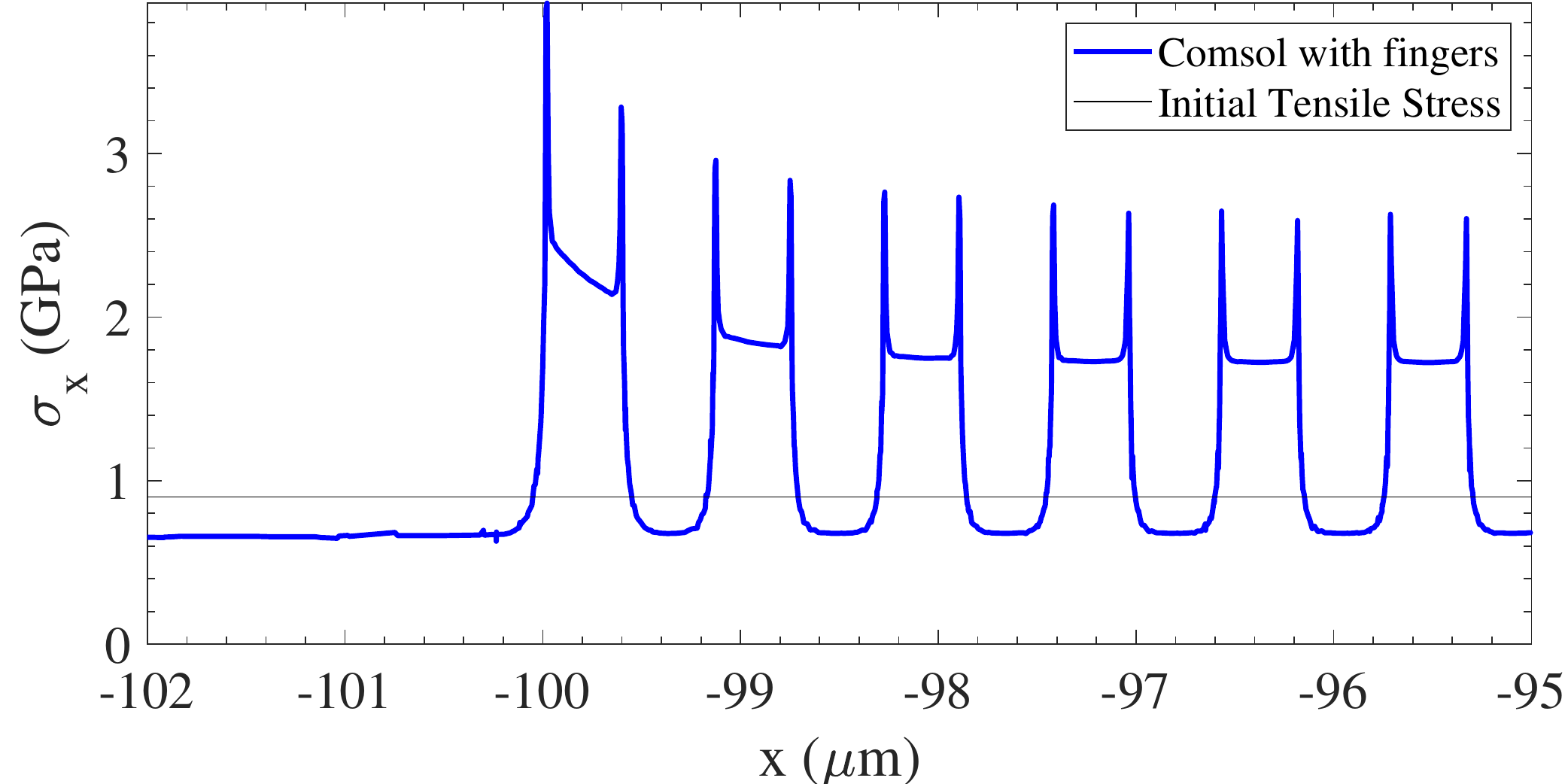}
\caption{Top: Full 3D simulations of the distribution of the stress component $\sigma_x$ along a transverse cut in the $x$-direction in the middle of sample A2 and for the first two grating periods of the etched area. Bottom: Variations in the $x$-direction of the simulated stress component $\sigma_x$ at $z=t/2-h$.}
\label{fig:stressG07}
\end{figure}

Figure~\ref{fig:stressG07} shows the simulated stress in the $x$-direction for $y=0$. The overall evolution observed in the one-dimensional model are confirmed; the stress, released in the unpatterned area, increases abruptly at the border of the patterned area before slowly decreasing towards the center of the structure. A noticeable difference with the 1D model, though, is due to the presence of the grating fingers and the periodic thickness modulation in the etched area. Abrupt increases and decreases in stress are thus observed at each finger edge, due to the periodic thickness changes. Let us note again that the sharp feature at each finger edge is quite dependent on the curvature of the trapezoidal corners in the simulations. Since, for the samples used in this work, the AFM tip angle and curvature do not allow for an precise determination of this curvature, we use a reasonable~\cite{Nair2019}, but somewhat arbitrary value of 20 nm in the simulations. The local stress values at the grating finger edges can thus only be considered as indicative under these conditions.

\subsubsection{Influence of finger height and SWG size}

\begin{figure*}[h!]
\centering
\includegraphics[width=\columnwidth]{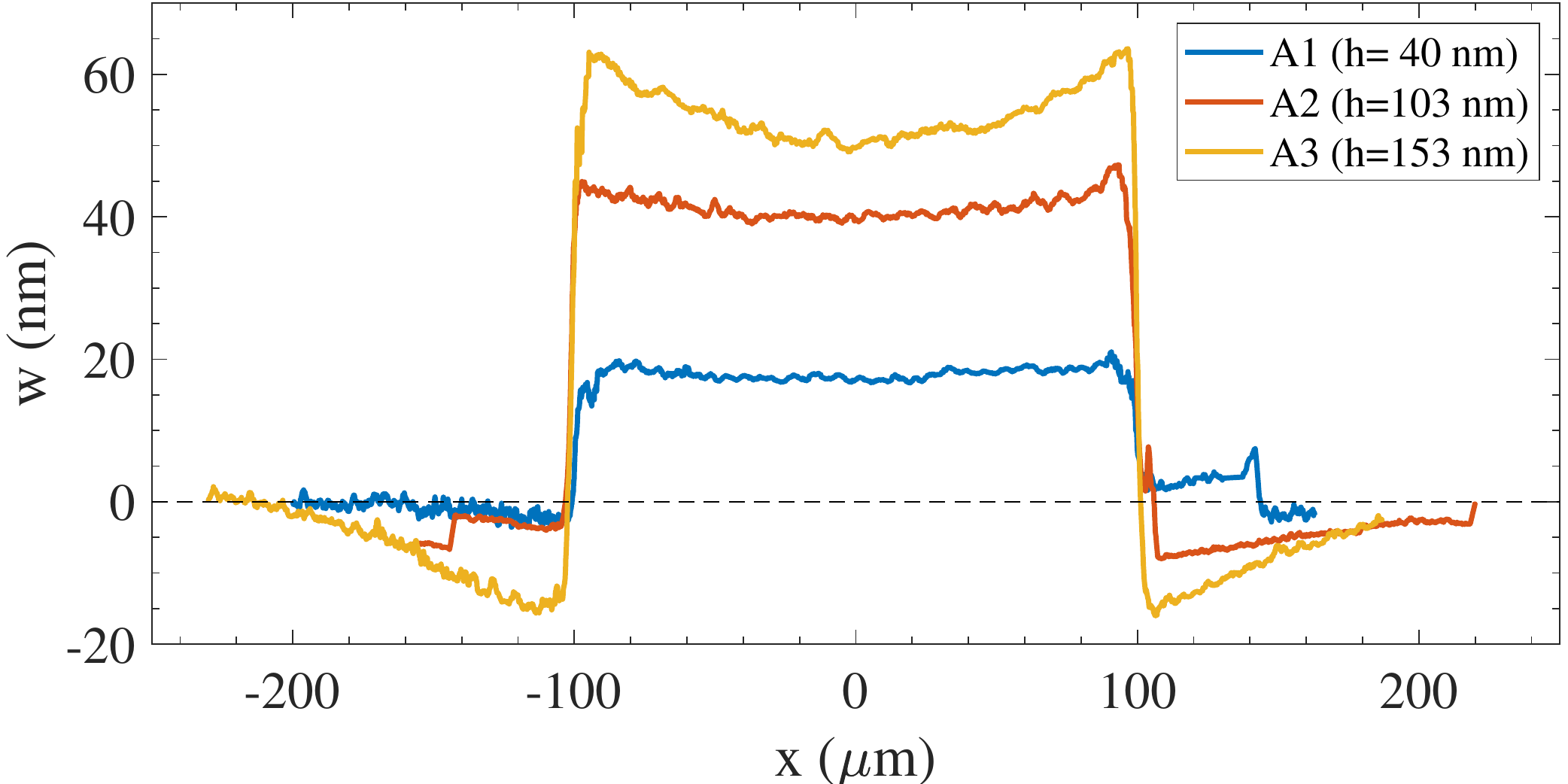}\includegraphics[width=\columnwidth]{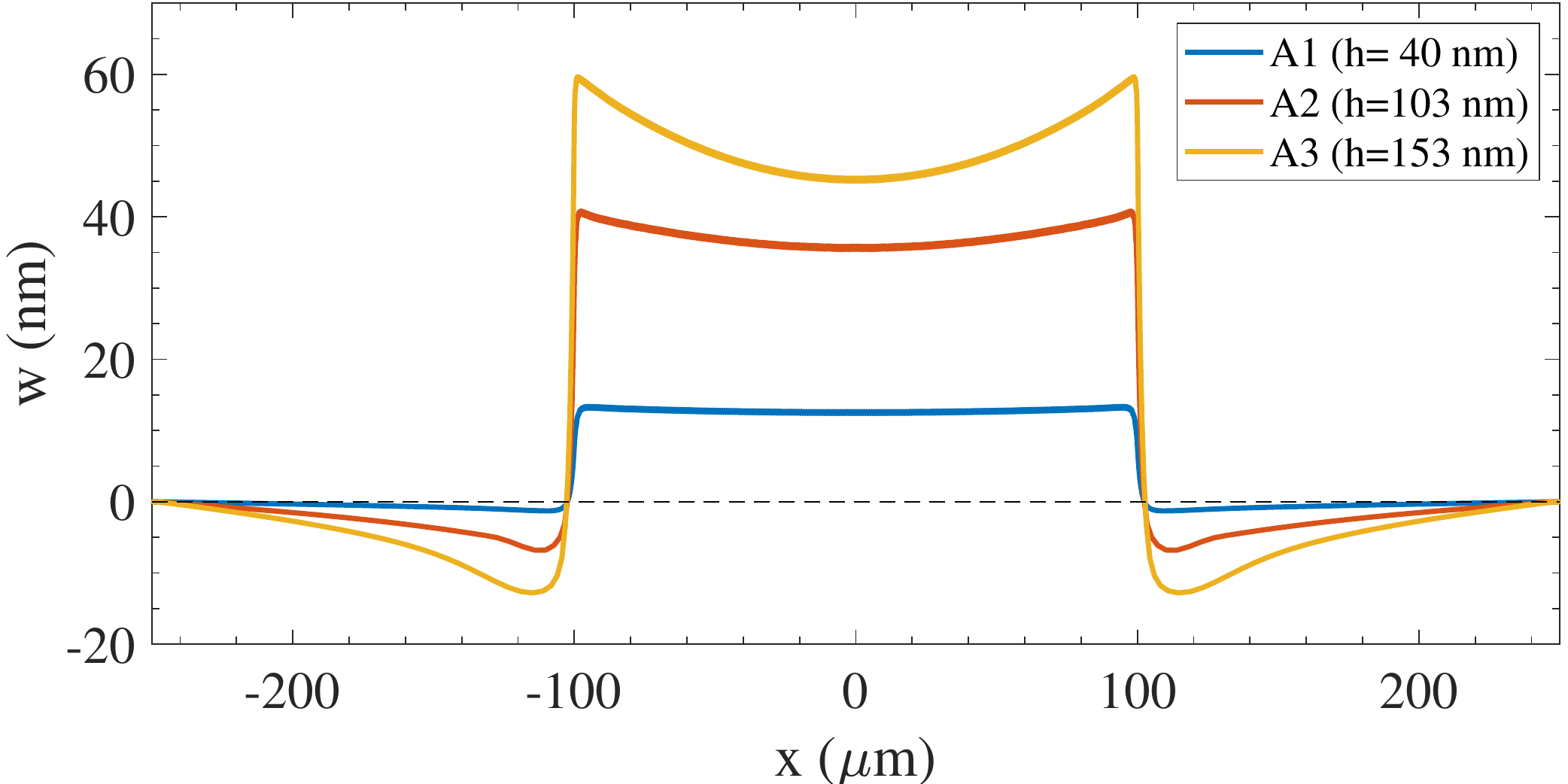}
\includegraphics[width=\columnwidth]{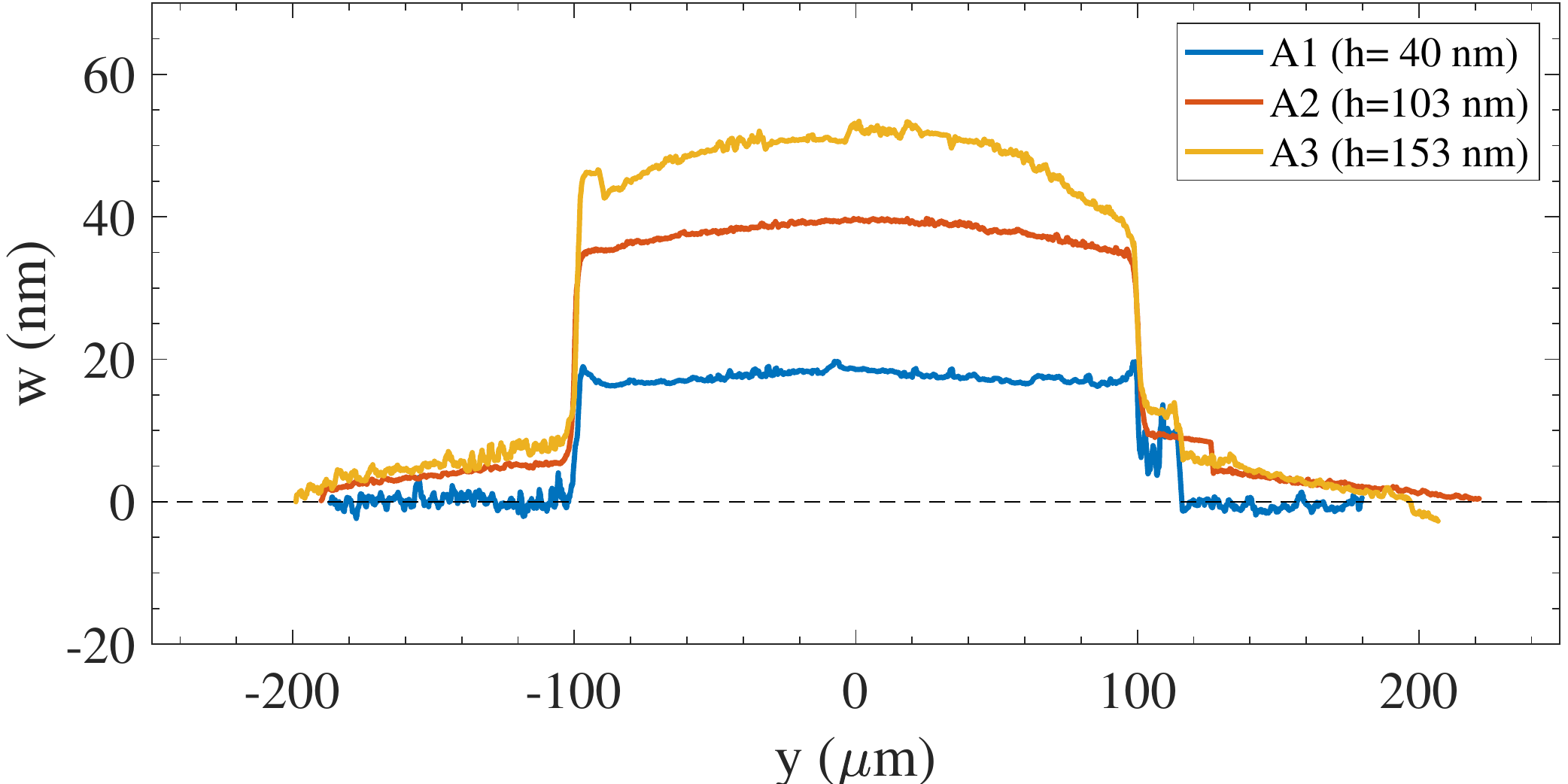}\includegraphics[width=\columnwidth]{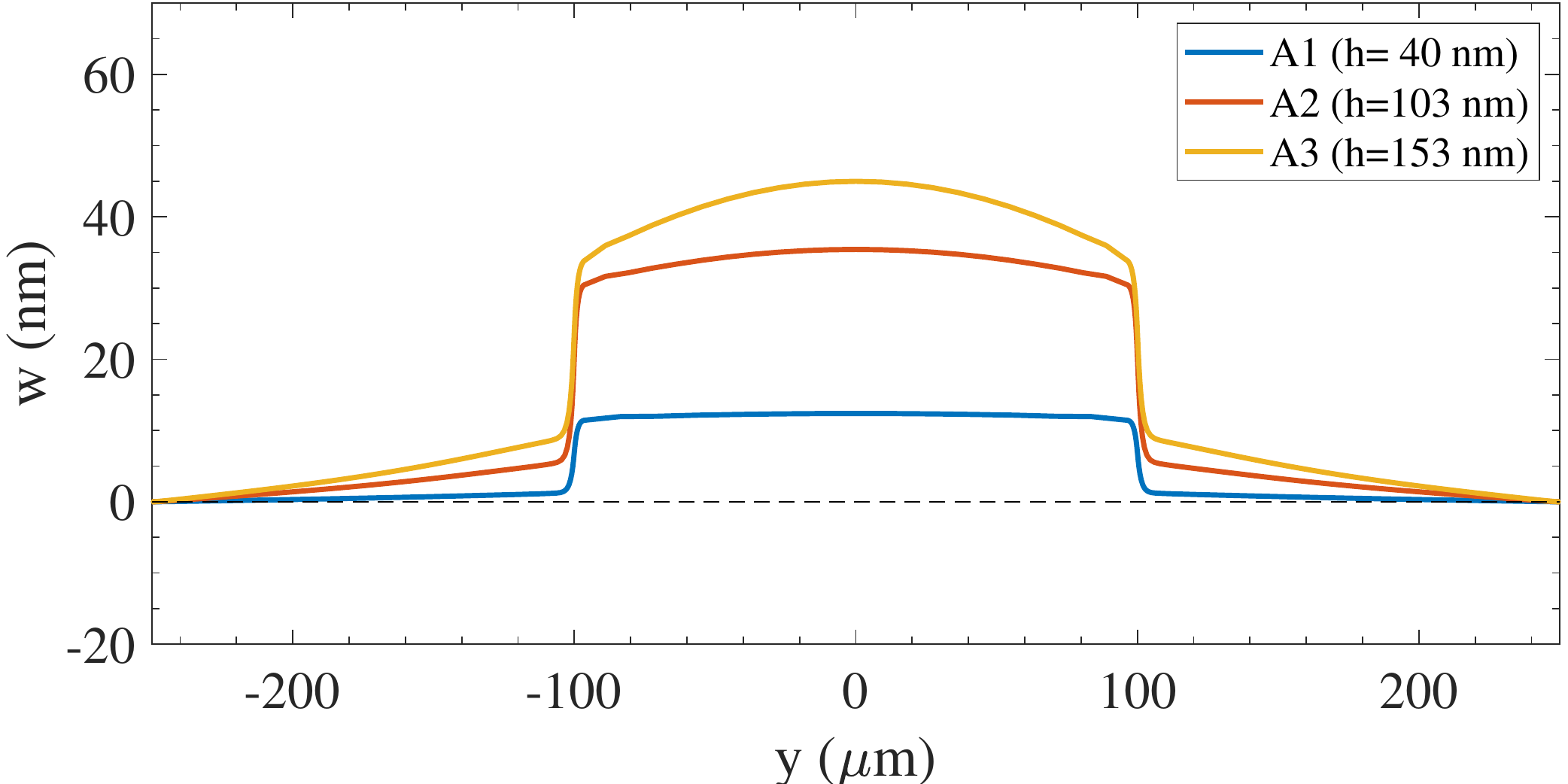}
\caption{Deflections studies for SWGs A1, A2, A3 with different finger heights $h=40$, 103 and 153 $\mu$m. Top and bottom left: measured deflections in the $x$- and $y$-directions. Top and bottom right: corresponding simulated deflections.}
\label{fig:depth}
\end{figure*}

\begin{figure*}[h!]
\centering
\includegraphics[width=\columnwidth]{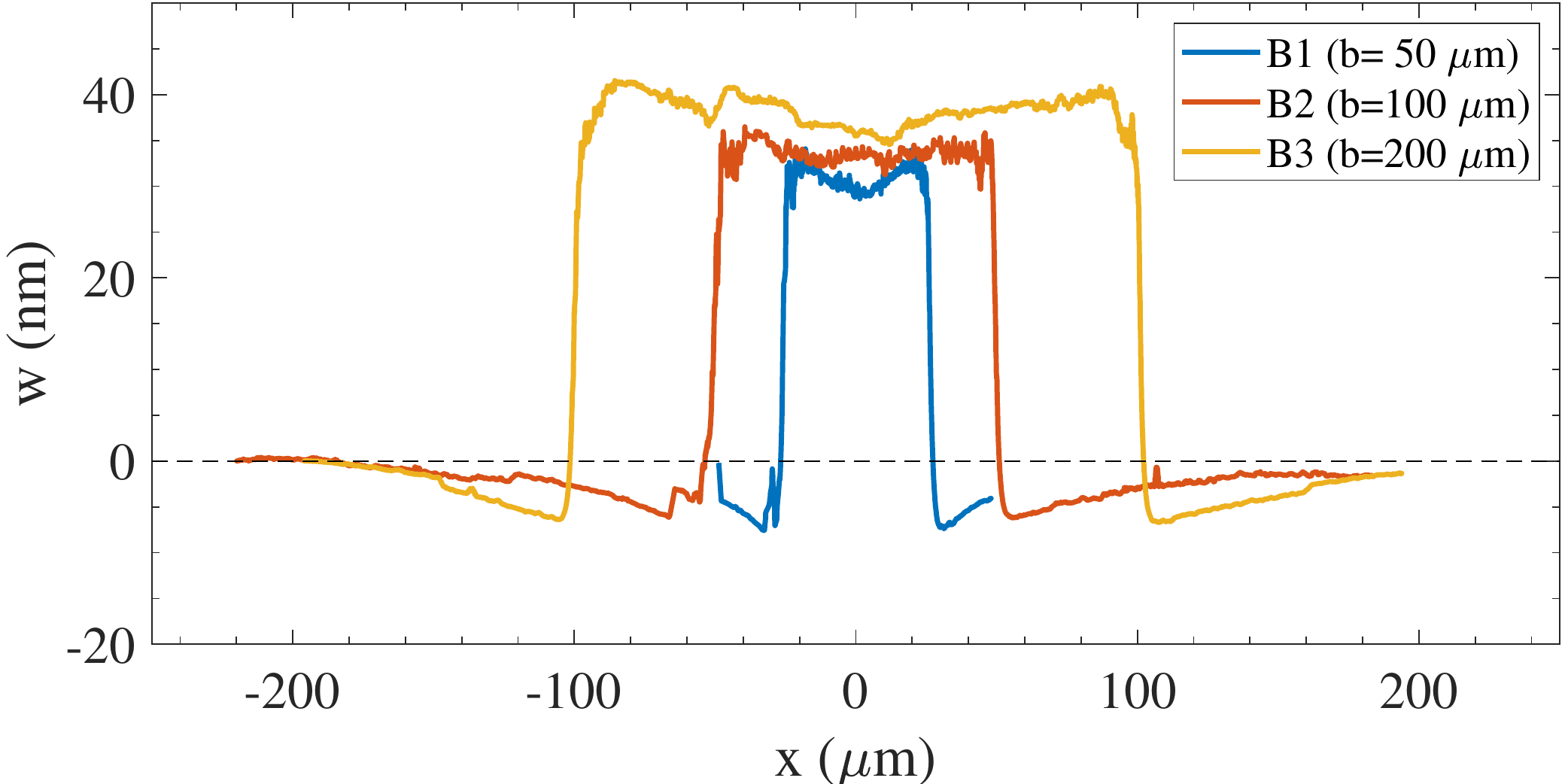}\includegraphics[width=\columnwidth]{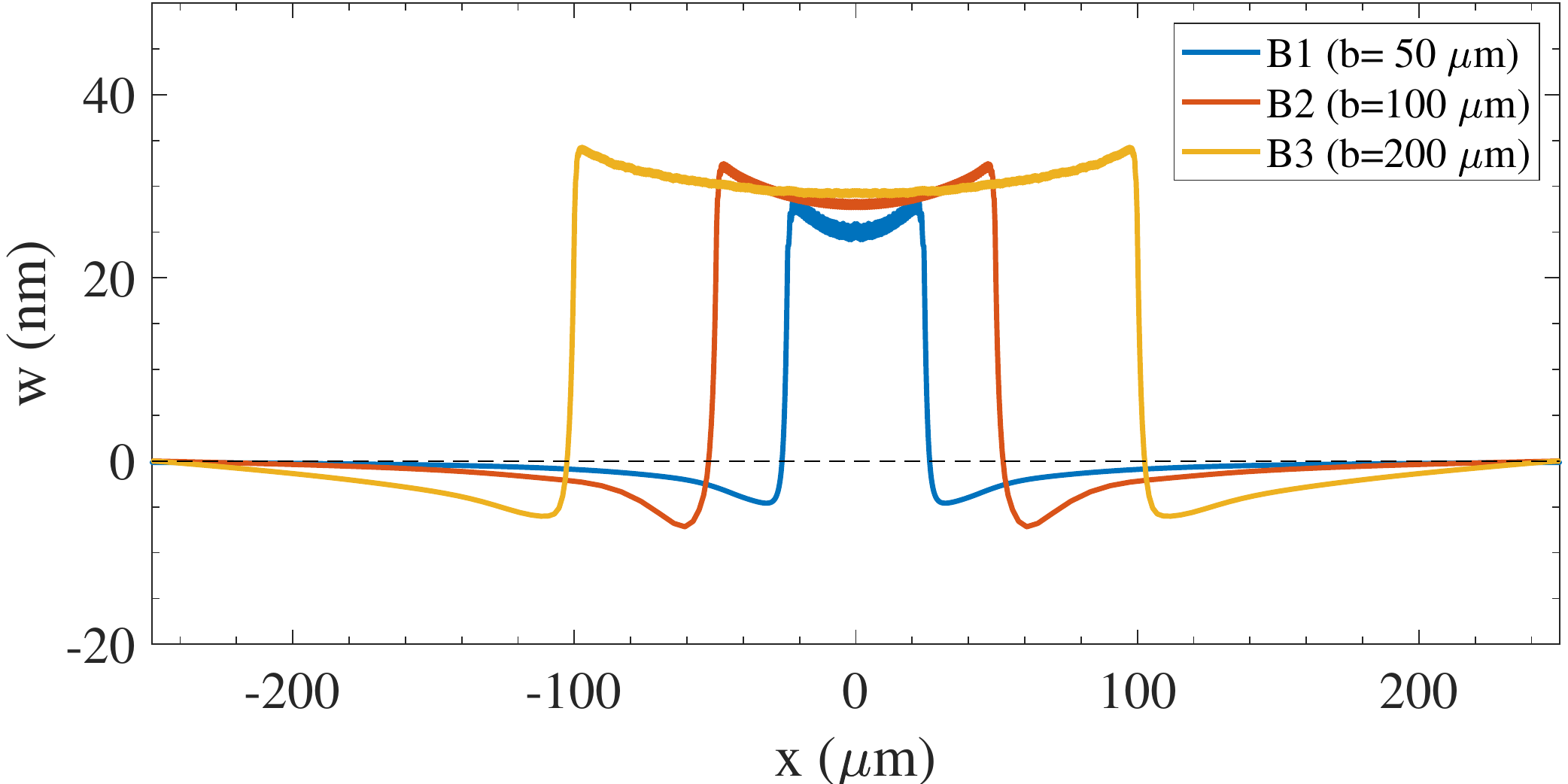}
\includegraphics[width=\columnwidth]{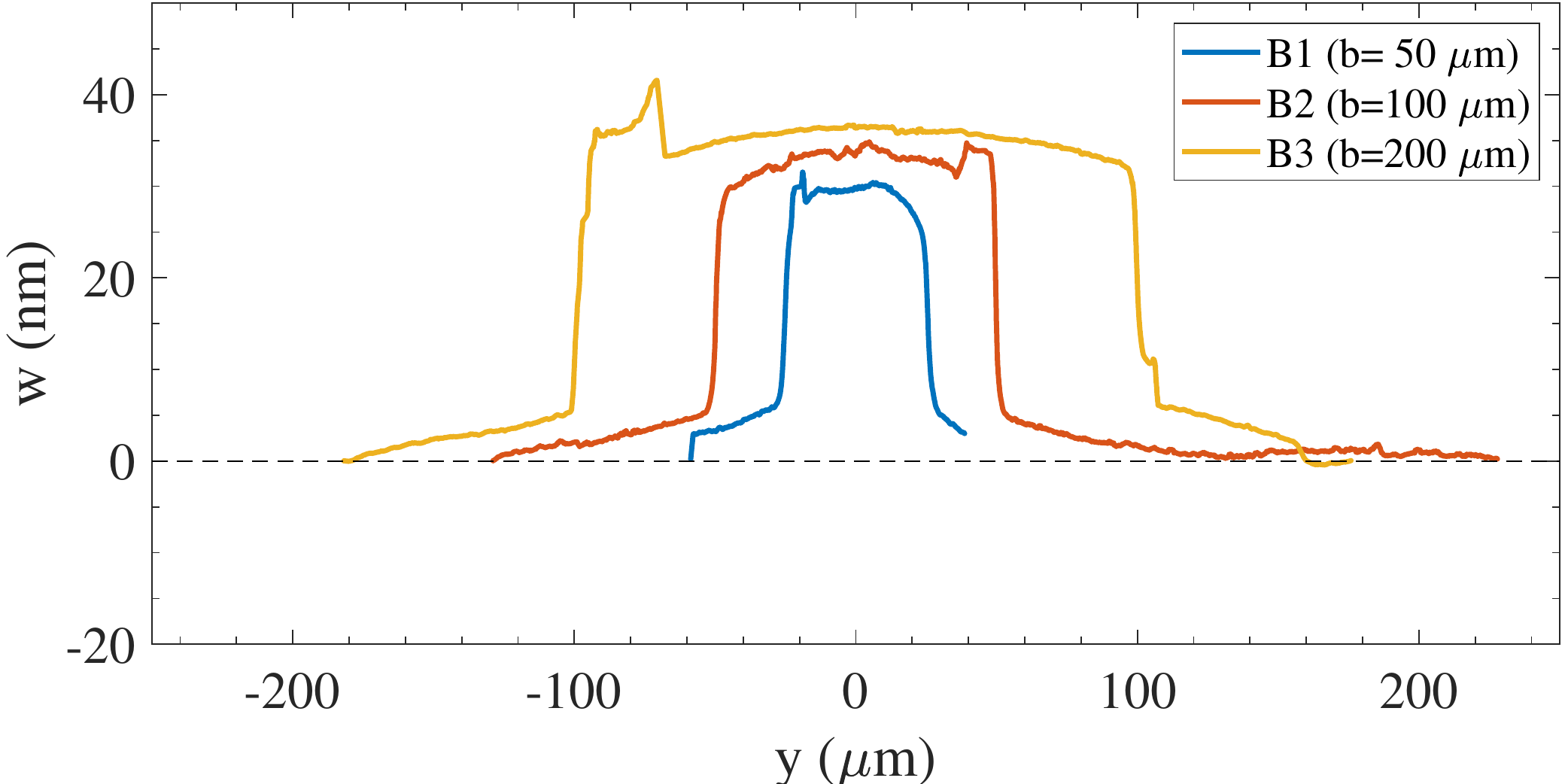}\includegraphics[width=\columnwidth]{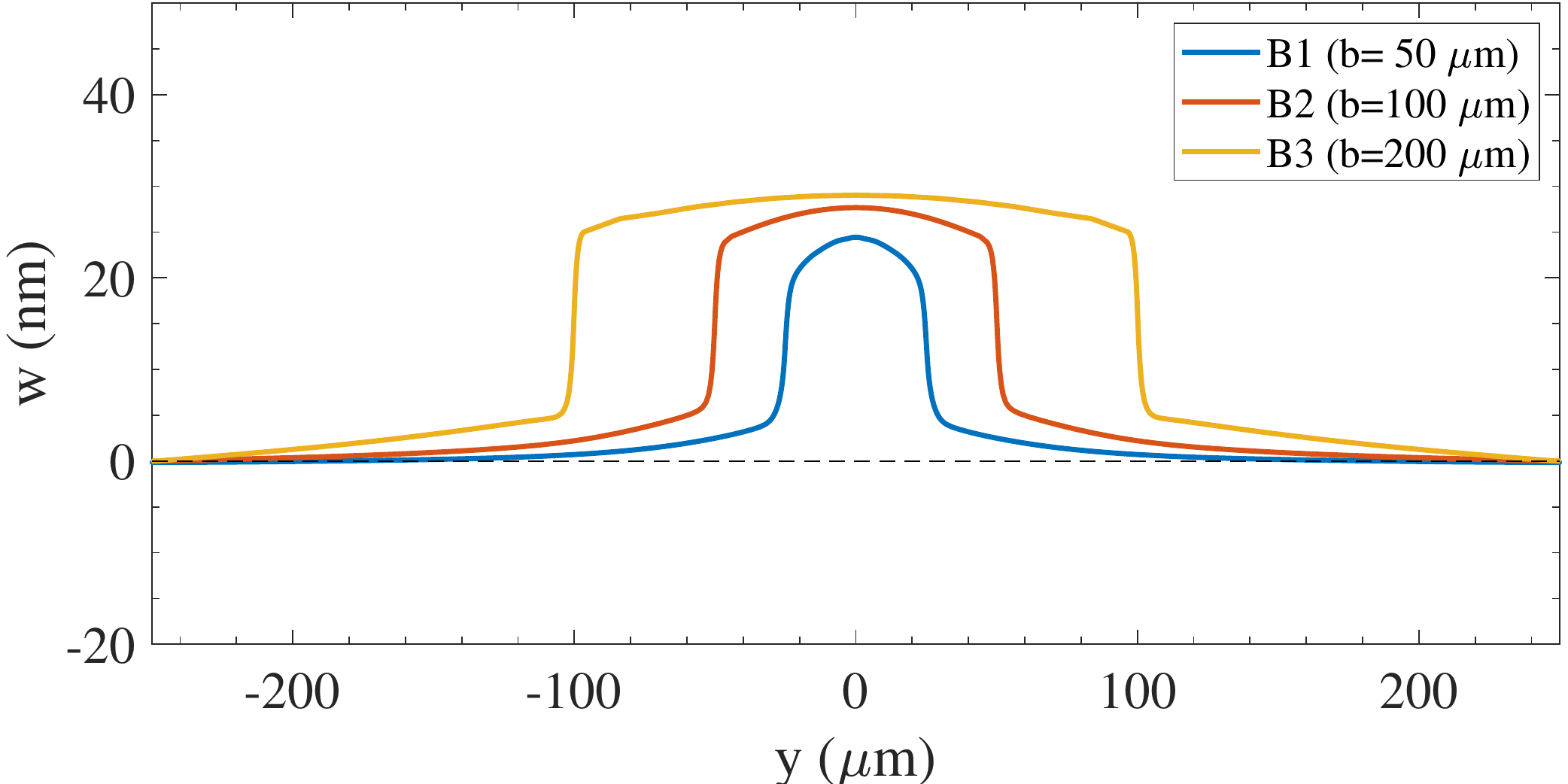}
\caption{Deflection studies for SWGs B1, B2, B3 with different sizes $b=50$, 100 and 200 $\mu$m. Top and bottom left: measured deflections in the $x$- and $y$-directions. Top and bottom right: corresponding simulated deflections.}
\label{fig:size}
\end{figure*}

Similar scans and simulations were performed for SWGs with different finger heights, as well as for SWGs with different sizes of the patterned area, all other parameters being similar (see Table~\ref{tab:parameters}). The results of long AFM scans in the $x$- and $y$-directions and crossing at the center of the SWGs are shown in Figs.~\ref{fig:depth} and \ref{fig:size}, respectively, together with the corresponding FEM simulation results. Baring the slight discrepancy between the final deflections level in the patterned area, the experimentally measured profiles globally agree well with the predictions of the simulations. 

When the etching depth is varied for a fixed SWG size and other grating parameters (Fig.~\ref{fig:depth}), a depression, which increases in amplitude as the depth increases, is observed in the $x$-direction at the outside of the patterned area, followed by an abrupt deflection leading to a maximum height change of approximately $h/2$ occuring over a few microns. A subsequent slowly decreasing deflection then occurs towards the center of the patterned area to match the maximum deflection in the $y$-direction. The negative and positive curvatures observed in the patterned area in the $x$- and $y$-directions, respectively, are also clearly seen to increase with the etching depth, as expected from the increased stress modifications. When the size of the SWG is increased for a fixed depth, Fig.~\ref{fig:size}) shows that the observed curvature decreases, as also expected from the model and the simulations.

\section{Conclusion}

In-depth investigations of the topology of freestanding thin silicon nitride films patterned with SWGs were carried out using AFM scans. This noninvasive profilometry method allows for extracting the relevant--and otherwise difficult to obtain--grating geometrical parameters, as well as to evidence the overall deflection of the films after the patterning process. We showed that the obtained grating parameters could be used as input to RCWA simulations to accurately predict the transmission spectrum of the SWGs under illumination with polarized light at normal incidence. The observed deflections of different membranes due to the modification of the tensile stress in the patterning process are well-captured by a simple model of a clamped membrane with varying thickness and are all in good agreement with the results of full three-dimensional FEM simulations. 

We believe that the observations and methods used in this work may be useful for the general understanding and design of nanostructured thin films, and in particular, may have implications for the nanomembranes patterned with photonic crystal or subwavelength gratings, which are currently applied in a number of applications within optomechanics and sensing. For instance, a consequence of these studies is that the strong increase in tensile stress at the edges of the patterned area may be expected to set a limit to the achievable etching depth by this fabrication method, as the fracture stress of the films may be reached. This will be the subject of future studies.

\begin{acknowledgments}
We acknowledge financial support from Independent Research Fund Denmark and the Danish Hydrocarbon research and Techology Center for finding software through project FL 15a, AWF.2.C.03. The data that support the findings of this study are available from the corresponding author upon request.
\end{acknowledgments}

\appendix*

\section{}

\begin{figure*}[h]
\centering
\includegraphics[width=\columnwidth]{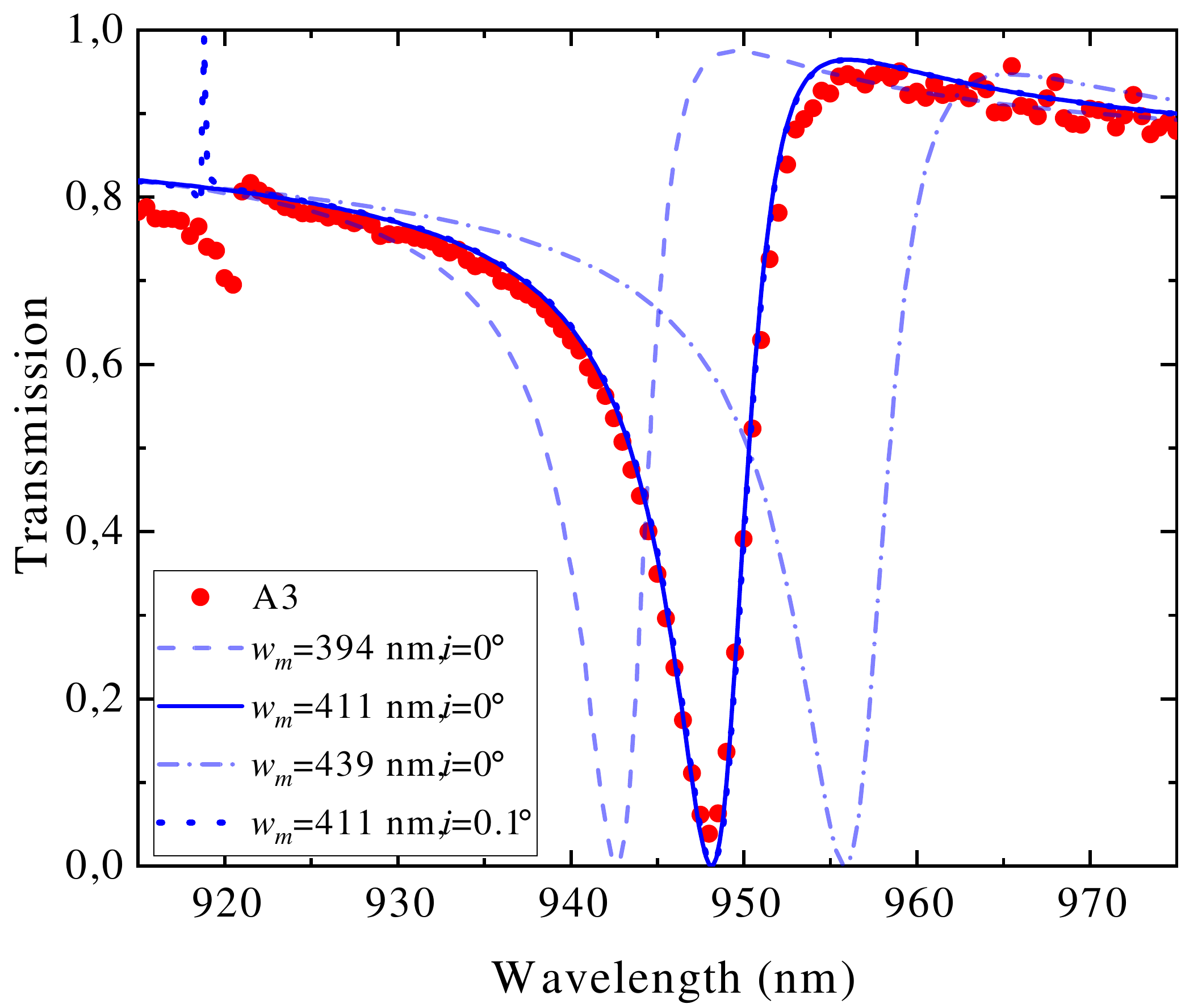}
\includegraphics[width=\columnwidth]{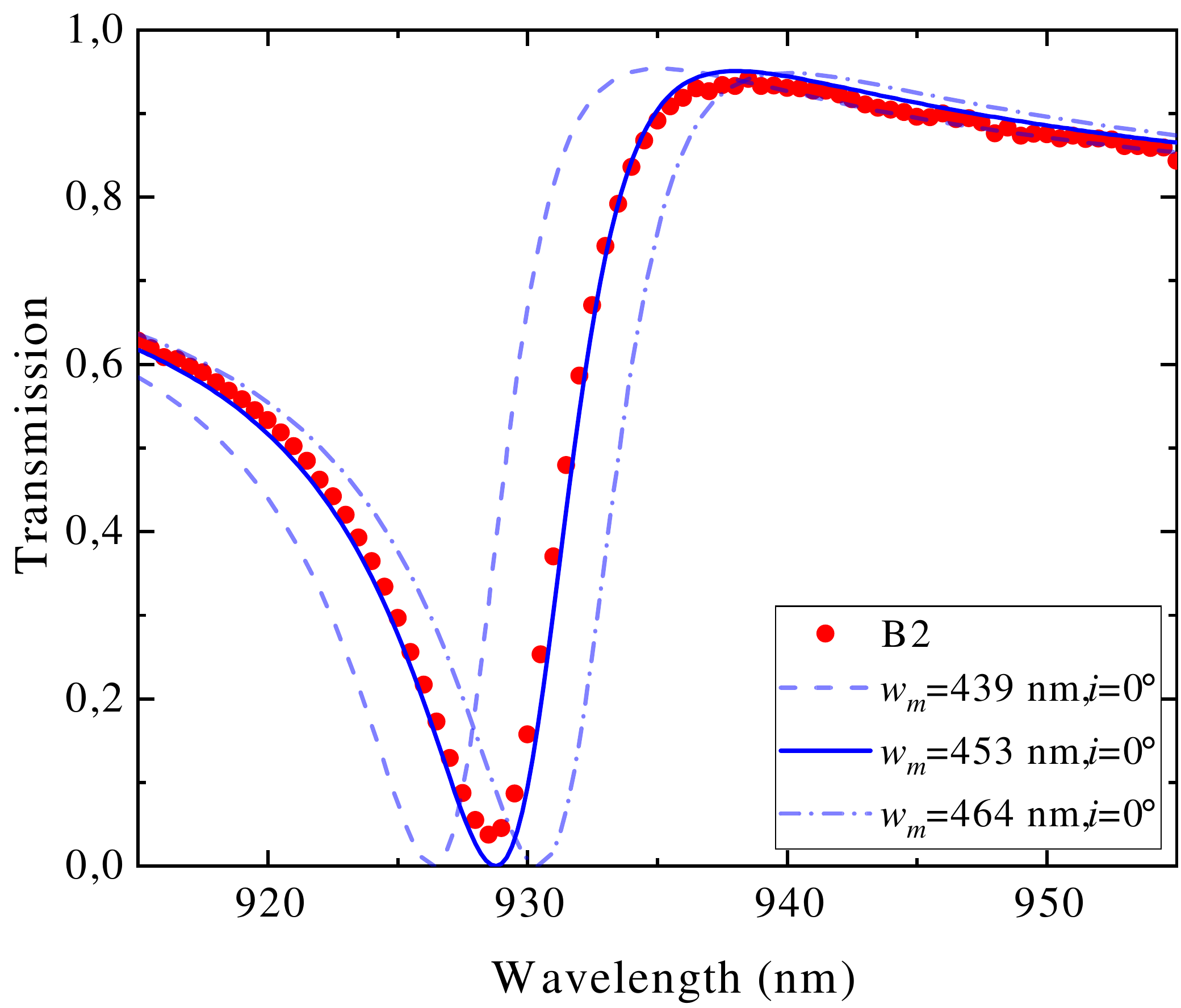}\\
\includegraphics[width=\columnwidth]{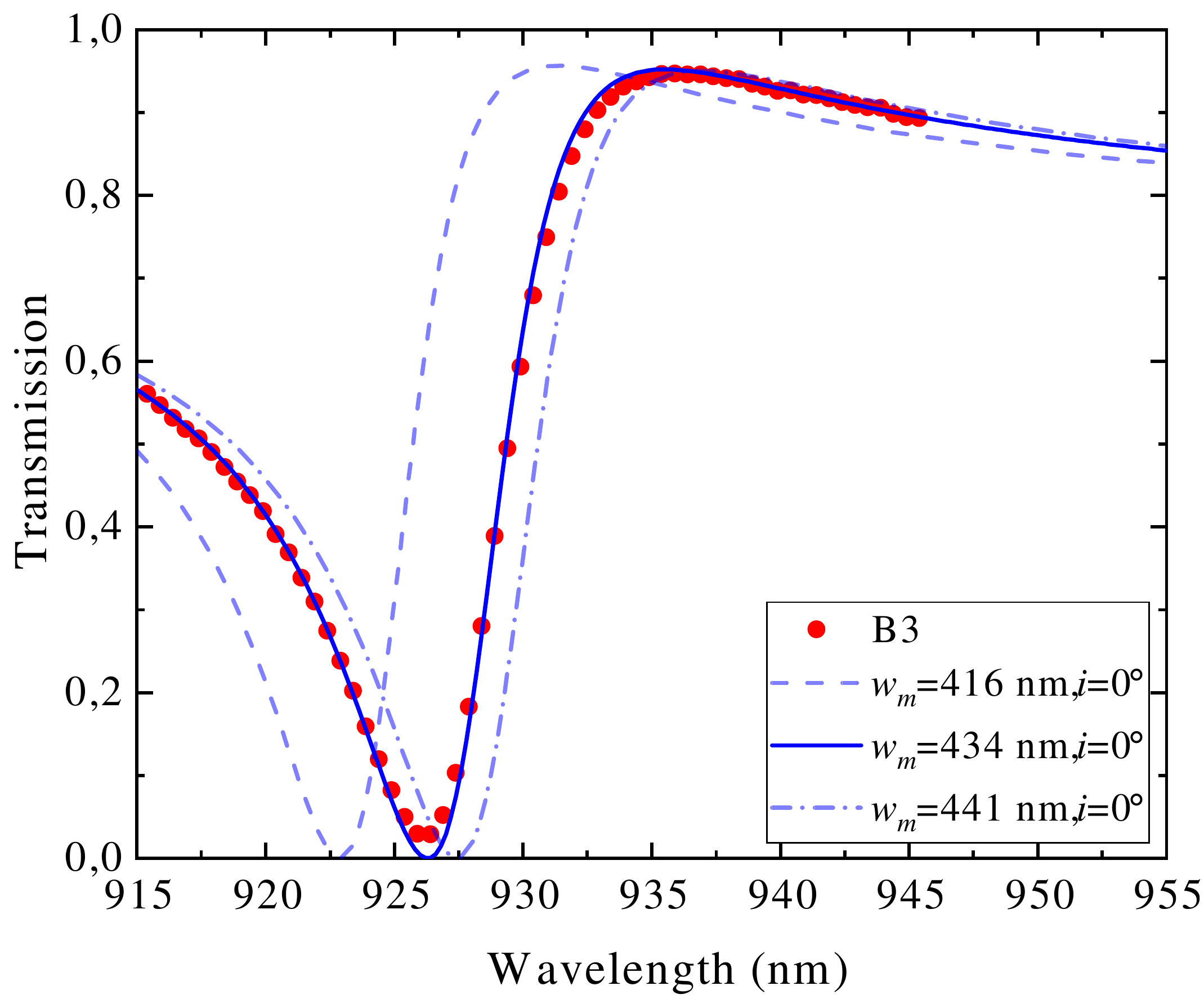}
\caption{Measured transmission spectra for samples A3, B2 and B3 (red dots), together with the RCWA simulated spectra for different mean finger widths/incidence angles (blue curves).}
\label{fig:appendix}
\end{figure*}


%

\clearpage
\bibliography{biblio_profilometry}

\begin{thebibliography}{53}%
\makeatletter
\providecommand \@ifxundefined [1]{%
 \@ifx{#1\undefined}
}%
\providecommand \@ifnum [1]{%
 \ifnum #1\expandafter \@firstoftwo
 \else \expandafter \@secondoftwo
 \fi
}%
\providecommand \@ifx [1]{%
 \ifx #1\expandafter \@firstoftwo
 \else \expandafter \@secondoftwo
 \fi
}%
\providecommand \natexlab [1]{#1}%
\providecommand \enquote  [1]{``#1''}%
\providecommand \bibnamefont  [1]{#1}%
\providecommand \bibfnamefont [1]{#1}%
\providecommand \citenamefont [1]{#1}%
\providecommand \href@noop [0]{\@secondoftwo}%
\providecommand \href [0]{\begingroup \@sanitize@url \@href}%
\providecommand \@href[1]{\@@startlink{#1}\@@href}%
\providecommand \@@href[1]{\endgroup#1\@@endlink}%
\providecommand \@sanitize@url [0]{\catcode `\\12\catcode `\$12\catcode
  `\&12\catcode `\#12\catcode `\^12\catcode `\_12\catcode `\%12\relax}%
\providecommand \@@startlink[1]{}%
\providecommand \@@endlink[0]{}%
\providecommand \url  [0]{\begingroup\@sanitize@url \@url }%
\providecommand \@url [1]{\endgroup\@href {#1}{\urlprefix }}%
\providecommand \urlprefix  [0]{URL }%
\providecommand \Eprint [0]{\href }%
\providecommand \doibase [0]{http://dx.doi.org/}%
\providecommand \selectlanguage [0]{\@gobble}%
\providecommand \bibinfo  [0]{\@secondoftwo}%
\providecommand \bibfield  [0]{\@secondoftwo}%
\providecommand \translation [1]{[#1]}%
\providecommand \BibitemOpen [0]{}%
\providecommand \bibitemStop [0]{}%
\providecommand \bibitemNoStop [0]{.\EOS\space}%
\providecommand \EOS [0]{\spacefactor3000\relax}%
\providecommand \BibitemShut  [1]{\csname bibitem#1\endcsname}%
\let\auto@bib@innerbib\@empty
\bibitem [{\citenamefont {Chang-Hasnain}\ and\ \citenamefont
  {Yang}(2012)}]{ChangHasnain2012}%
  \BibitemOpen
  \bibfield  {author} {\bibinfo {author} {\bibfnamefont {C.~J.}\ \bibnamefont
  {Chang-Hasnain}}\ and\ \bibinfo {author} {\bibfnamefont {W.}~\bibnamefont
  {Yang}},\ }\bibfield  {title} {\enquote {\bibinfo {title} {High-contrast
  gratings for integrated optoelectronics},}\ }\href {\doibase
  10.1364/AOP.4.000379} {\bibfield  {journal} {\bibinfo  {journal} {Adv. Opt.
  Photon.}\ }\textbf {\bibinfo {volume} {4}},\ \bibinfo {pages} {379--440}
  (\bibinfo {year} {2012})}\BibitemShut {NoStop}%
\bibitem [{\citenamefont {Zhou}\ \emph {et~al.}(2014)\citenamefont {Zhou},
  \citenamefont {Zhao}, \citenamefont {Shuai}, \citenamefont {Yang},
  \citenamefont {Chuwongin}, \citenamefont {Chadha}, \citenamefont {Seo},
  \citenamefont {Wang}, \citenamefont {Liu}, \citenamefont {Ma},\ and\
  \citenamefont {Fan}}]{Zhou2014}%
  \BibitemOpen
  \bibfield  {author} {\bibinfo {author} {\bibfnamefont {W.}~\bibnamefont
  {Zhou}}, \bibinfo {author} {\bibfnamefont {D.}~\bibnamefont {Zhao}}, \bibinfo
  {author} {\bibfnamefont {Y.-C.}\ \bibnamefont {Shuai}}, \bibinfo {author}
  {\bibfnamefont {H.}~\bibnamefont {Yang}}, \bibinfo {author} {\bibfnamefont
  {S.}~\bibnamefont {Chuwongin}}, \bibinfo {author} {\bibfnamefont
  {A.}~\bibnamefont {Chadha}}, \bibinfo {author} {\bibfnamefont {J.-H.}\
  \bibnamefont {Seo}}, \bibinfo {author} {\bibfnamefont {K.~X.}\ \bibnamefont
  {Wang}}, \bibinfo {author} {\bibfnamefont {V.}~\bibnamefont {Liu}}, \bibinfo
  {author} {\bibfnamefont {Z.}~\bibnamefont {Ma}}, \ and\ \bibinfo {author}
  {\bibfnamefont {S.}~\bibnamefont {Fan}},\ }\bibfield  {title} {\enquote
  {\bibinfo {title} {Progress in 2d photonic crystal fano resonance
  photonics},}\ }\href {\doibase
  https://doi.org/10.1016/j.pquantelec.2014.01.001} {\bibfield  {journal}
  {\bibinfo  {journal} {Progress in Quantum Electronics}\ }\textbf {\bibinfo
  {volume} {38}},\ \bibinfo {pages} {1 -- 74} (\bibinfo {year}
  {2014})}\BibitemShut {NoStop}%
\bibitem [{\citenamefont {Quaranta}\ \emph {et~al.}()\citenamefont {Quaranta},
  \citenamefont {Basset}, \citenamefont {Martin},\ and\ \citenamefont
  {Gallinet}}]{Quaranta2018}%
  \BibitemOpen
  \bibfield  {author} {\bibinfo {author} {\bibfnamefont {G.}~\bibnamefont
  {Quaranta}}, \bibinfo {author} {\bibfnamefont {G.}~\bibnamefont {Basset}},
  \bibinfo {author} {\bibfnamefont {O.~J.~F.}\ \bibnamefont {Martin}}, \ and\
  \bibinfo {author} {\bibfnamefont {B.}~\bibnamefont {Gallinet}},\ }\bibfield
  {title} {\enquote {\bibinfo {title} {Recent advances in resonant waveguide
  gratings},}\ }\href {\doibase 10.1002/lpor.201800017} {\bibfield  {journal}
  {\bibinfo  {journal} {Laser \& Photonics Reviews}\ }\textbf {\bibinfo
  {volume} {12}},\ \bibinfo {pages} {1800017}},\ \Eprint
  {http://arxiv.org/abs/https://onlinelibrary.wiley.com/doi/pdf/10.1002/lpor.201800017}
  {https://onlinelibrary.wiley.com/doi/pdf/10.1002/lpor.201800017} \BibitemShut
  {NoStop}%
\bibitem [{\citenamefont {Cheben}\ \emph {et~al.}(2018)\citenamefont {Cheben},
  \citenamefont {Halir}, \citenamefont {Schmid}, \citenamefont {Atwater},\ and\
  \citenamefont {Smith}}]{Cheben2018}%
  \BibitemOpen
  \bibfield  {author} {\bibinfo {author} {\bibfnamefont {P.}~\bibnamefont
  {Cheben}}, \bibinfo {author} {\bibfnamefont {R.}~\bibnamefont {Halir}},
  \bibinfo {author} {\bibfnamefont {J.~H.}\ \bibnamefont {Schmid}}, \bibinfo
  {author} {\bibfnamefont {H.~A.}\ \bibnamefont {Atwater}}, \ and\ \bibinfo
  {author} {\bibfnamefont {D.~R.}\ \bibnamefont {Smith}},\ }\bibfield  {title}
  {\enquote {\bibinfo {title} {Subwavelength integrated photonics},}\ }\href
  {\doibase 10.1038/s41586-018-0421-7} {\bibfield  {journal} {\bibinfo
  {journal} {Nature}\ }\textbf {\bibinfo {volume} {560}},\ \bibinfo {pages}
  {565--572} (\bibinfo {year} {2018})}\BibitemShut {NoStop}%
\bibitem [{\citenamefont {Shuai}\ \emph {et~al.}(2013)\citenamefont {Shuai},
  \citenamefont {Zhao}, \citenamefont {Tian}, \citenamefont {Seo},
  \citenamefont {Plant}, \citenamefont {Ma}, \citenamefont {Fan},\ and\
  \citenamefont {Zhou}}]{Shuai2013}%
  \BibitemOpen
  \bibfield  {author} {\bibinfo {author} {\bibfnamefont {Y.}~\bibnamefont
  {Shuai}}, \bibinfo {author} {\bibfnamefont {D.}~\bibnamefont {Zhao}},
  \bibinfo {author} {\bibfnamefont {Z.}~\bibnamefont {Tian}}, \bibinfo {author}
  {\bibfnamefont {J.-H.}\ \bibnamefont {Seo}}, \bibinfo {author} {\bibfnamefont
  {D.~V.}\ \bibnamefont {Plant}}, \bibinfo {author} {\bibfnamefont
  {Z.}~\bibnamefont {Ma}}, \bibinfo {author} {\bibfnamefont {S.}~\bibnamefont
  {Fan}}, \ and\ \bibinfo {author} {\bibfnamefont {W.}~\bibnamefont {Zhou}},\
  }\bibfield  {title} {\enquote {\bibinfo {title} {Double-layer fano resonance
  photonic crystal filters},}\ }\href {\doibase 10.1364/OE.21.024582}
  {\bibfield  {journal} {\bibinfo  {journal} {Opt. Express}\ }\textbf {\bibinfo
  {volume} {21}},\ \bibinfo {pages} {24582--24589} (\bibinfo {year}
  {2013})}\BibitemShut {NoStop}%
\bibitem [{\citenamefont {Wang}\ \emph {et~al.}(2015)\citenamefont {Wang},
  \citenamefont {Stellinga}, \citenamefont {Klemm}, \citenamefont {Reardon},\
  and\ \citenamefont {Krauss}}]{Wang2015}%
  \BibitemOpen
  \bibfield  {author} {\bibinfo {author} {\bibfnamefont {Y.}~\bibnamefont
  {Wang}}, \bibinfo {author} {\bibfnamefont {D.}~\bibnamefont {Stellinga}},
  \bibinfo {author} {\bibfnamefont {A.~B.}\ \bibnamefont {Klemm}}, \bibinfo
  {author} {\bibfnamefont {C.~P.}\ \bibnamefont {Reardon}}, \ and\ \bibinfo
  {author} {\bibfnamefont {T.~F.}\ \bibnamefont {Krauss}},\ }\bibfield  {title}
  {\enquote {\bibinfo {title} {Tunable optical filters based on silicon nitride
  high contrast gratings},}\ }\href {\doibase 10.1109/JSTQE.2014.2377656}
  {\bibfield  {journal} {\bibinfo  {journal} {IEEE Journal of Selected Topics
  in Quantum Electronics}\ }\textbf {\bibinfo {volume} {21}},\ \bibinfo {pages}
  {108--113} (\bibinfo {year} {2015})}\BibitemShut {NoStop}%
\bibitem [{\citenamefont {Zhou}, \citenamefont {Huang},\ and\ \citenamefont
  {Chang-Hasnain}(2008)}]{Zhou2008}%
  \BibitemOpen
  \bibfield  {author} {\bibinfo {author} {\bibfnamefont {Y.}~\bibnamefont
  {Zhou}}, \bibinfo {author} {\bibfnamefont {M.~C.~Y.}\ \bibnamefont {Huang}},
  \ and\ \bibinfo {author} {\bibfnamefont {C.~J.}\ \bibnamefont
  {Chang-Hasnain}},\ }\bibfield  {title} {\enquote {\bibinfo {title} {Tunable
  vcsel with ultra-thin high contrast grating for high-speed tuning},}\ }\href
  {\doibase 10.1364/OE.16.014221} {\bibfield  {journal} {\bibinfo  {journal}
  {Opt. Express}\ }\textbf {\bibinfo {volume} {16}},\ \bibinfo {pages}
  {14221--14226} (\bibinfo {year} {2008})}\BibitemShut {NoStop}%
\bibitem [{\citenamefont {Penades}\ \emph {et~al.}(2016)\citenamefont
  {Penades}, \citenamefont {{n}ux}, \citenamefont {Nedeljkovic}, \citenamefont
  {Wang\"{u}emert-P\'{e}rez}, \citenamefont {Halir}, \citenamefont {Khokhar},
  \citenamefont {Alonso-Ramos}, \citenamefont {Qu}, \citenamefont
  {Molina-Fern\'{a}ndez}, \citenamefont {Cheben},\ and\ \citenamefont
  {Mashanovich}}]{Penades2016}%
  \BibitemOpen
  \bibfield  {author} {\bibinfo {author} {\bibfnamefont {J.~S.}\ \bibnamefont
  {Penades}}, \bibinfo {author} {\bibfnamefont {A.~O.-M.}\ \bibnamefont
  {{n}ux}}, \bibinfo {author} {\bibfnamefont {M.}~\bibnamefont {Nedeljkovic}},
  \bibinfo {author} {\bibfnamefont {J.~G.}\ \bibnamefont
  {Wang\"{u}emert-P\'{e}rez}}, \bibinfo {author} {\bibfnamefont
  {R.}~\bibnamefont {Halir}}, \bibinfo {author} {\bibfnamefont {A.~Z.}\
  \bibnamefont {Khokhar}}, \bibinfo {author} {\bibfnamefont {C.}~\bibnamefont
  {Alonso-Ramos}}, \bibinfo {author} {\bibfnamefont {Z.}~\bibnamefont {Qu}},
  \bibinfo {author} {\bibfnamefont {I.}~\bibnamefont {Molina-Fern\'{a}ndez}},
  \bibinfo {author} {\bibfnamefont {P.}~\bibnamefont {Cheben}}, \ and\ \bibinfo
  {author} {\bibfnamefont {G.~Z.}\ \bibnamefont {Mashanovich}},\ }\bibfield
  {title} {\enquote {\bibinfo {title} {Suspended silicon mid-infrared waveguide
  devices with subwavelength grating metamaterial cladding},}\ }\href {\doibase
  10.1364/OE.24.022908} {\bibfield  {journal} {\bibinfo  {journal} {Opt.
  Express}\ }\textbf {\bibinfo {volume} {24}},\ \bibinfo {pages} {22908--22916}
  (\bibinfo {year} {2016})}\BibitemShut {NoStop}%
\bibitem [{\citenamefont {Kang}\ \emph {et~al.}(2017)\citenamefont {Kang},
  \citenamefont {Cheng}, \citenamefont {Zhou}, \citenamefont {Xiao},
  \citenamefont {Gopalakrisna}, \citenamefont {Takenaka}, \citenamefont
  {Tsang},\ and\ \citenamefont {Goda}}]{Kang2017}%
  \BibitemOpen
  \bibfield  {author} {\bibinfo {author} {\bibfnamefont {J.}~\bibnamefont
  {Kang}}, \bibinfo {author} {\bibfnamefont {Z.}~\bibnamefont {Cheng}},
  \bibinfo {author} {\bibfnamefont {W.}~\bibnamefont {Zhou}}, \bibinfo {author}
  {\bibfnamefont {T.-H.}\ \bibnamefont {Xiao}}, \bibinfo {author}
  {\bibfnamefont {K.-L.}\ \bibnamefont {Gopalakrisna}}, \bibinfo {author}
  {\bibfnamefont {M.}~\bibnamefont {Takenaka}}, \bibinfo {author}
  {\bibfnamefont {H.~K.}\ \bibnamefont {Tsang}}, \ and\ \bibinfo {author}
  {\bibfnamefont {K.}~\bibnamefont {Goda}},\ }\bibfield  {title} {\enquote
  {\bibinfo {title} {Focusing subwavelength grating coupler for mid-infrared
  suspended membrane germanium waveguides},}\ }\href {\doibase
  10.1364/OL.42.002094} {\bibfield  {journal} {\bibinfo  {journal} {Opt.
  Lett.}\ }\textbf {\bibinfo {volume} {42}},\ \bibinfo {pages} {2094--2097}
  (\bibinfo {year} {2017})}\BibitemShut {NoStop}%
\bibitem [{\citenamefont {Br\"uckner}\ \emph {et~al.}(2010)\citenamefont
  {Br\"uckner}, \citenamefont {Friedrich}, \citenamefont {Clausnitzer},
  \citenamefont {Britzger}, \citenamefont {Burmeister}, \citenamefont
  {Danzmann}, \citenamefont {Kley}, \citenamefont {T\"unnermann},\ and\
  \citenamefont {Schnabel}}]{Bruckner2010}%
  \BibitemOpen
  \bibfield  {author} {\bibinfo {author} {\bibfnamefont {F.}~\bibnamefont
  {Br\"uckner}}, \bibinfo {author} {\bibfnamefont {D.}~\bibnamefont
  {Friedrich}}, \bibinfo {author} {\bibfnamefont {T.}~\bibnamefont
  {Clausnitzer}}, \bibinfo {author} {\bibfnamefont {M.}~\bibnamefont
  {Britzger}}, \bibinfo {author} {\bibfnamefont {O.}~\bibnamefont
  {Burmeister}}, \bibinfo {author} {\bibfnamefont {K.}~\bibnamefont
  {Danzmann}}, \bibinfo {author} {\bibfnamefont {E.-B.}\ \bibnamefont {Kley}},
  \bibinfo {author} {\bibfnamefont {A.}~\bibnamefont {T\"unnermann}}, \ and\
  \bibinfo {author} {\bibfnamefont {R.}~\bibnamefont {Schnabel}},\ }\bibfield
  {title} {\enquote {\bibinfo {title} {Realization of a monolithic
  high-reflectivity cavity mirror from a single silicon crystal},}\ }\href
  {\doibase 10.1103/PhysRevLett.104.163903} {\bibfield  {journal} {\bibinfo
  {journal} {Phys. Rev. Lett.}\ }\textbf {\bibinfo {volume} {104}},\ \bibinfo
  {pages} {163903} (\bibinfo {year} {2010})}\BibitemShut {NoStop}%
\bibitem [{\citenamefont {Fattal}\ \emph {et~al.}(2010)\citenamefont {Fattal},
  \citenamefont {Li}, \citenamefont {Peng}, \citenamefont {Fiorentino},\ and\
  \citenamefont {Beausoleil}}]{Fattal2010}%
  \BibitemOpen
  \bibfield  {author} {\bibinfo {author} {\bibfnamefont {D.}~\bibnamefont
  {Fattal}}, \bibinfo {author} {\bibfnamefont {J.}~\bibnamefont {Li}}, \bibinfo
  {author} {\bibfnamefont {Z.}~\bibnamefont {Peng}}, \bibinfo {author}
  {\bibfnamefont {M.}~\bibnamefont {Fiorentino}}, \ and\ \bibinfo {author}
  {\bibfnamefont {R.~G.}\ \bibnamefont {Beausoleil}},\ }\bibfield  {title}
  {\enquote {\bibinfo {title} {Flat dielectric grating reflectors with focusing
  abilities},}\ }\href {http://dx.doi.org/10.1038/nphoton.2010.116} {\bibfield
  {journal} {\bibinfo  {journal} {Nature Photonics}\ }\textbf {\bibinfo
  {volume} {4}},\ \bibinfo {pages} {466} (\bibinfo {year} {2010})}\BibitemShut
  {NoStop}%
\bibitem [{\citenamefont {Lu}\ \emph {et~al.}(2010)\citenamefont {Lu},
  \citenamefont {Sedgwick}, \citenamefont {Karagodsky}, \citenamefont {Chase},\
  and\ \citenamefont {Chang-Hasnain}}]{Lu2010}%
  \BibitemOpen
  \bibfield  {author} {\bibinfo {author} {\bibfnamefont {F.}~\bibnamefont
  {Lu}}, \bibinfo {author} {\bibfnamefont {F.~G.}\ \bibnamefont {Sedgwick}},
  \bibinfo {author} {\bibfnamefont {V.}~\bibnamefont {Karagodsky}}, \bibinfo
  {author} {\bibfnamefont {C.}~\bibnamefont {Chase}}, \ and\ \bibinfo {author}
  {\bibfnamefont {C.~J.}\ \bibnamefont {Chang-Hasnain}},\ }\bibfield  {title}
  {\enquote {\bibinfo {title} {Planar high-numerical-aperture low-loss focusing
  reflectors and lenses using subwavelength high contrast gratings},}\ }\href
  {\doibase 10.1364/OE.18.012606} {\bibfield  {journal} {\bibinfo  {journal}
  {Opt. Express}\ }\textbf {\bibinfo {volume} {18}},\ \bibinfo {pages}
  {12606--12614} (\bibinfo {year} {2010})}\BibitemShut {NoStop}%
\bibitem [{\citenamefont {Klemm}\ \emph {et~al.}(2013)\citenamefont {Klemm},
  \citenamefont {Stellinga}, \citenamefont {Martins}, \citenamefont {Lewis},
  \citenamefont {Huyet}, \citenamefont {O'Faolain},\ and\ \citenamefont
  {Krauss}}]{Klemm2013}%
  \BibitemOpen
  \bibfield  {author} {\bibinfo {author} {\bibfnamefont {A.~B.}\ \bibnamefont
  {Klemm}}, \bibinfo {author} {\bibfnamefont {D.}~\bibnamefont {Stellinga}},
  \bibinfo {author} {\bibfnamefont {E.~R.}\ \bibnamefont {Martins}}, \bibinfo
  {author} {\bibfnamefont {L.}~\bibnamefont {Lewis}}, \bibinfo {author}
  {\bibfnamefont {G.}~\bibnamefont {Huyet}}, \bibinfo {author} {\bibfnamefont
  {L.}~\bibnamefont {O'Faolain}}, \ and\ \bibinfo {author} {\bibfnamefont
  {T.~F.}\ \bibnamefont {Krauss}},\ }\bibfield  {title} {\enquote {\bibinfo
  {title} {Experimental high numerical aperture focusing with high contrast
  gratings},}\ }\href {\doibase 10.1364/OL.38.003410} {\bibfield  {journal}
  {\bibinfo  {journal} {Opt. Lett.}\ }\textbf {\bibinfo {volume} {38}},\
  \bibinfo {pages} {3410--3413} (\bibinfo {year} {2013})}\BibitemShut {NoStop}%
\bibitem [{\citenamefont {Mutlu}\ \emph {et~al.}(2012)\citenamefont {Mutlu},
  \citenamefont {Akosman}, \citenamefont {Kurt}, \citenamefont {Gokkavas},\
  and\ \citenamefont {Ozbay}}]{Mutlu2012}%
  \BibitemOpen
  \bibfield  {author} {\bibinfo {author} {\bibfnamefont {M.}~\bibnamefont
  {Mutlu}}, \bibinfo {author} {\bibfnamefont {A.~E.}\ \bibnamefont {Akosman}},
  \bibinfo {author} {\bibfnamefont {G.}~\bibnamefont {Kurt}}, \bibinfo {author}
  {\bibfnamefont {M.}~\bibnamefont {Gokkavas}}, \ and\ \bibinfo {author}
  {\bibfnamefont {E.}~\bibnamefont {Ozbay}},\ }\bibfield  {title} {\enquote
  {\bibinfo {title} {Experimental realization of a high-contrast grating based
  broadband quarter-wave plate},}\ }\href {\doibase 10.1364/OE.20.027966}
  {\bibfield  {journal} {\bibinfo  {journal} {Opt. Express}\ }\textbf {\bibinfo
  {volume} {20}},\ \bibinfo {pages} {27966--27973} (\bibinfo {year}
  {2012})}\BibitemShut {NoStop}%
\bibitem [{\citenamefont {Bykov}\ \emph {et~al.}(2018)\citenamefont {Bykov},
  \citenamefont {Doskolovich}, \citenamefont {Morozov}, \citenamefont
  {Podlipnov}, \citenamefont {Bezus}, \citenamefont {Verma},\ and\
  \citenamefont {Soifer}}]{Bykov2018}%
  \BibitemOpen
  \bibfield  {author} {\bibinfo {author} {\bibfnamefont {D.~A.}\ \bibnamefont
  {Bykov}}, \bibinfo {author} {\bibfnamefont {L.~L.}\ \bibnamefont
  {Doskolovich}}, \bibinfo {author} {\bibfnamefont {A.~A.}\ \bibnamefont
  {Morozov}}, \bibinfo {author} {\bibfnamefont {V.~V.}\ \bibnamefont
  {Podlipnov}}, \bibinfo {author} {\bibfnamefont {E.~A.}\ \bibnamefont
  {Bezus}}, \bibinfo {author} {\bibfnamefont {P.}~\bibnamefont {Verma}}, \ and\
  \bibinfo {author} {\bibfnamefont {V.~A.}\ \bibnamefont {Soifer}},\ }\bibfield
   {title} {\enquote {\bibinfo {title} {First-order optical spatial
  differentiator based on a guided-mode resonant grating},}\ }\href {\doibase
  10.1364/OE.26.010997} {\bibfield  {journal} {\bibinfo  {journal} {Opt.
  Express}\ }\textbf {\bibinfo {volume} {26}},\ \bibinfo {pages} {10997--11006}
  (\bibinfo {year} {2018})}\BibitemShut {NoStop}%
\bibitem [{\citenamefont {Dong}\ \emph {et~al.}(2018)\citenamefont {Dong},
  \citenamefont {Si}, \citenamefont {Yu},\ and\ \citenamefont
  {Deng}}]{Dong2018}%
  \BibitemOpen
  \bibfield  {author} {\bibinfo {author} {\bibfnamefont {Z.}~\bibnamefont
  {Dong}}, \bibinfo {author} {\bibfnamefont {J.}~\bibnamefont {Si}}, \bibinfo
  {author} {\bibfnamefont {X.}~\bibnamefont {Yu}}, \ and\ \bibinfo {author}
  {\bibfnamefont {X.}~\bibnamefont {Deng}},\ }\bibfield  {title} {\enquote
  {\bibinfo {title} {Optical spatial differentiator based on subwavelength
  high-contrast gratings},}\ }\href {\doibase 10.1063/1.5026309} {\bibfield
  {journal} {\bibinfo  {journal} {Applied Physics Letters}\ }\textbf {\bibinfo
  {volume} {112}},\ \bibinfo {pages} {181102} (\bibinfo {year} {2018})},\
  \Eprint {http://arxiv.org/abs/https://doi.org/10.1063/1.5026309}
  {https://doi.org/10.1063/1.5026309} \BibitemShut {NoStop}%
\bibitem [{\citenamefont {Yang}\ \emph {et~al.}(2020)\citenamefont {Yang},
  \citenamefont {Yu}, \citenamefont {Zhang},\ and\ \citenamefont
  {Deng}}]{Yang2020}%
  \BibitemOpen
  \bibfield  {author} {\bibinfo {author} {\bibfnamefont {W.}~\bibnamefont
  {Yang}}, \bibinfo {author} {\bibfnamefont {X.}~\bibnamefont {Yu}}, \bibinfo
  {author} {\bibfnamefont {J.}~\bibnamefont {Zhang}}, \ and\ \bibinfo {author}
  {\bibfnamefont {X.}~\bibnamefont {Deng}},\ }\bibfield  {title} {\enquote
  {\bibinfo {title} {Plasmonic transmitted optical differentiator based on the
  subwavelength gold gratings},}\ }\href {\doibase 10.1364/OL.390566}
  {\bibfield  {journal} {\bibinfo  {journal} {Opt. Lett.}\ }\textbf {\bibinfo
  {volume} {45}},\ \bibinfo {pages} {2295--2298} (\bibinfo {year}
  {2020})}\BibitemShut {NoStop}%
\bibitem [{\citenamefont {Parthenopoulos}\ \emph {et~al.}(2020)\citenamefont
  {Parthenopoulos}, \citenamefont {Darki}, \citenamefont {Jeppesen},\ and\
  \citenamefont {Dantan}}]{Parthenopoulos2020}%
  \BibitemOpen
  \bibfield  {author} {\bibinfo {author} {\bibfnamefont {A.}~\bibnamefont
  {Parthenopoulos}}, \bibinfo {author} {\bibfnamefont {A.~A.}\ \bibnamefont
  {Darki}}, \bibinfo {author} {\bibfnamefont {B.~R.}\ \bibnamefont {Jeppesen}},
  \ and\ \bibinfo {author} {\bibfnamefont {A.}~\bibnamefont {Dantan}},\
  }\bibfield  {title} {\enquote {\bibinfo {title} {Optical spatial
  differentiation using suspended subwavelength gratings},}\ }\href@noop {}
  {\bibfield  {journal} {\bibinfo  {journal} {arxiv:2008.10945}\ } (\bibinfo
  {year} {2020})}\BibitemShut {NoStop}%
\bibitem [{\citenamefont {Boutami}\ \emph {et~al.}(2007)\citenamefont
  {Boutami}, \citenamefont {Benbakir}, \citenamefont {Letartre}, \citenamefont
  {Leclercq}, \citenamefont {Regreny},\ and\ \citenamefont
  {Viktorovitch}}]{Boutami2007}%
  \BibitemOpen
  \bibfield  {author} {\bibinfo {author} {\bibfnamefont {S.}~\bibnamefont
  {Boutami}}, \bibinfo {author} {\bibfnamefont {B.}~\bibnamefont {Benbakir}},
  \bibinfo {author} {\bibfnamefont {X.}~\bibnamefont {Letartre}}, \bibinfo
  {author} {\bibfnamefont {J.}~\bibnamefont {Leclercq}}, \bibinfo {author}
  {\bibfnamefont {P.}~\bibnamefont {Regreny}}, \ and\ \bibinfo {author}
  {\bibfnamefont {P.}~\bibnamefont {Viktorovitch}},\ }\bibfield  {title}
  {\enquote {\bibinfo {title} {Ultimate vertical fabry-perot cavity based on
  single-layer photonic crystal mirrors},}\ }\href {\doibase
  10.1364/OE.15.012443} {\bibfield  {journal} {\bibinfo  {journal} {Opt.
  Express}\ }\textbf {\bibinfo {volume} {15}},\ \bibinfo {pages} {12443--12449}
  (\bibinfo {year} {2007})}\BibitemShut {NoStop}%
\bibitem [{\citenamefont {Huang}, \citenamefont {Zhou},\ and\ \citenamefont
  {Chang-Hasnain}(2008)}]{Huang2008}%
  \BibitemOpen
  \bibfield  {author} {\bibinfo {author} {\bibfnamefont {M.~C.~Y.}\
  \bibnamefont {Huang}}, \bibinfo {author} {\bibfnamefont {Y.}~\bibnamefont
  {Zhou}}, \ and\ \bibinfo {author} {\bibfnamefont {C.~J.}\ \bibnamefont
  {Chang-Hasnain}},\ }\bibfield  {title} {\enquote {\bibinfo {title} {A
  nanoelectromechanical tunable laser},}\ }\href
  {http://dx.doi.org/10.1038/nphoton.2008.3} {\bibfield  {journal} {\bibinfo
  {journal} {Nature Photonics}\ }\textbf {\bibinfo {volume} {2}},\ \bibinfo
  {pages} {180} (\bibinfo {year} {2008})}\BibitemShut {NoStop}%
\bibitem [{\citenamefont {Wagner}\ \emph {et~al.}(2016)\citenamefont {Wagner},
  \citenamefont {Sudzius}, \citenamefont {Mischok}, \citenamefont {Fröb},\
  and\ \citenamefont {Leo}}]{Wagner2016}%
  \BibitemOpen
  \bibfield  {author} {\bibinfo {author} {\bibfnamefont {T.}~\bibnamefont
  {Wagner}}, \bibinfo {author} {\bibfnamefont {M.}~\bibnamefont {Sudzius}},
  \bibinfo {author} {\bibfnamefont {A.}~\bibnamefont {Mischok}}, \bibinfo
  {author} {\bibfnamefont {H.}~\bibnamefont {Fröb}}, \ and\ \bibinfo {author}
  {\bibfnamefont {K.}~\bibnamefont {Leo}},\ }\bibfield  {title} {\enquote
  {\bibinfo {title} {Cross-coupled composite-cavity organic microresonators},}\
  }\href {\doibase 10.1063/1.4960095} {\bibfield  {journal} {\bibinfo
  {journal} {Applied Physics Letters}\ }\textbf {\bibinfo {volume} {109}},\
  \bibinfo {pages} {043302} (\bibinfo {year} {2016})},\ \Eprint
  {http://arxiv.org/abs/https://doi.org/10.1063/1.4960095}
  {https://doi.org/10.1063/1.4960095} \BibitemShut {NoStop}%
\bibitem [{\citenamefont {Midolo}, \citenamefont {Schliesser},\ and\
  \citenamefont {Fiore}(2018)}]{Midolo2018}%
  \BibitemOpen
  \bibfield  {author} {\bibinfo {author} {\bibfnamefont {L.}~\bibnamefont
  {Midolo}}, \bibinfo {author} {\bibfnamefont {A.}~\bibnamefont {Schliesser}},
  \ and\ \bibinfo {author} {\bibfnamefont {A.}~\bibnamefont {Fiore}},\
  }\bibfield  {title} {\enquote {\bibinfo {title} {Nano-opto-electro-mechanical
  systems},}\ }\href {\doibase 10.1038/s41565-017-0039-1} {\bibfield  {journal}
  {\bibinfo  {journal} {Nature Nanotechnology}\ }\textbf {\bibinfo {volume}
  {13}},\ \bibinfo {pages} {11--18} (\bibinfo {year} {2018})}\BibitemShut
  {NoStop}%
\bibitem [{\citenamefont {Ierardi}\ \emph {et~al.}(2014)\citenamefont
  {Ierardi}, \citenamefont {Becker}, \citenamefont {Pantazis}, \citenamefont
  {Firpo}, \citenamefont {Valbusa},\ and\ \citenamefont
  {Jousten}}]{Ierardi2014}%
  \BibitemOpen
  \bibfield  {author} {\bibinfo {author} {\bibfnamefont {V.}~\bibnamefont
  {Ierardi}}, \bibinfo {author} {\bibfnamefont {U.}~\bibnamefont {Becker}},
  \bibinfo {author} {\bibfnamefont {S.}~\bibnamefont {Pantazis}}, \bibinfo
  {author} {\bibfnamefont {G.}~\bibnamefont {Firpo}}, \bibinfo {author}
  {\bibfnamefont {U.}~\bibnamefont {Valbusa}}, \ and\ \bibinfo {author}
  {\bibfnamefont {K.}~\bibnamefont {Jousten}},\ }\bibfield  {title} {\enquote
  {\bibinfo {title} {Nano-holes as standard leak elements},}\ }\href {\doibase
  https://doi.org/10.1016/j.measurement.2014.09.017} {\bibfield  {journal}
  {\bibinfo  {journal} {Measurement}\ }\textbf {\bibinfo {volume} {58}},\
  \bibinfo {pages} {335 -- 341} (\bibinfo {year} {2014})}\BibitemShut {NoStop}%
\bibitem [{\citenamefont {Peltonen}\ \emph {et~al.}(2016)\citenamefont
  {Peltonen}, \citenamefont {Nguyen}, \citenamefont {Muhonen},\ and\
  \citenamefont {Pekola}}]{Peltonen2016}%
  \BibitemOpen
  \bibfield  {author} {\bibinfo {author} {\bibfnamefont {A.}~\bibnamefont
  {Peltonen}}, \bibinfo {author} {\bibfnamefont {H.~Q.}\ \bibnamefont
  {Nguyen}}, \bibinfo {author} {\bibfnamefont {J.~T.}\ \bibnamefont {Muhonen}},
  \ and\ \bibinfo {author} {\bibfnamefont {J.~P.}\ \bibnamefont {Pekola}},\
  }\bibfield  {title} {\enquote {\bibinfo {title} {Milling a silicon nitride
  membrane by focused ion beam},}\ }\href {\doibase 10.1116/1.4963895}
  {\bibfield  {journal} {\bibinfo  {journal} {Journal of Vacuum Science \&
  Technology B}\ }\textbf {\bibinfo {volume} {34}},\ \bibinfo {pages} {062201}
  (\bibinfo {year} {2016})},\ \Eprint
  {http://arxiv.org/abs/https://doi.org/10.1116/1.4963895}
  {https://doi.org/10.1116/1.4963895} \BibitemShut {NoStop}%
\bibitem [{\citenamefont {Nair}\ \emph {et~al.}(2019)\citenamefont {Nair},
  \citenamefont {Naesby}, \citenamefont {Jeppesen},\ and\ \citenamefont
  {Dantan}}]{Nair2019}%
  \BibitemOpen
  \bibfield  {author} {\bibinfo {author} {\bibfnamefont {B.}~\bibnamefont
  {Nair}}, \bibinfo {author} {\bibfnamefont {A.}~\bibnamefont {Naesby}},
  \bibinfo {author} {\bibfnamefont {B.~R.}\ \bibnamefont {Jeppesen}}, \ and\
  \bibinfo {author} {\bibfnamefont {A.}~\bibnamefont {Dantan}},\ }\bibfield
  {title} {\enquote {\bibinfo {title} {Suspended silicon nitride thin films
  with enhanced and electrically tunable reflectivity},}\ }\href {\doibase
  10.1088/1402-4896/ab3d6f} {\bibfield  {journal} {\bibinfo  {journal} {Physica
  Scripta}\ }\textbf {\bibinfo {volume} {94}},\ \bibinfo {pages} {125013}
  (\bibinfo {year} {2019})}\BibitemShut {NoStop}%
\bibitem [{\citenamefont {{Chang-Hasnain}}\ \emph {et~al.}(2009)\citenamefont
  {{Chang-Hasnain}}, \citenamefont {{Zhou}}, \citenamefont {{Huang}},\ and\
  \citenamefont {{Chase}}}]{ChangHasnain2009}%
  \BibitemOpen
  \bibfield  {author} {\bibinfo {author} {\bibfnamefont {C.~J.}\ \bibnamefont
  {{Chang-Hasnain}}}, \bibinfo {author} {\bibfnamefont {Y.}~\bibnamefont
  {{Zhou}}}, \bibinfo {author} {\bibfnamefont {M.~C.~Y.}\ \bibnamefont
  {{Huang}}}, \ and\ \bibinfo {author} {\bibfnamefont {C.}~\bibnamefont
  {{Chase}}},\ }\bibfield  {title} {\enquote {\bibinfo {title} {High-contrast
  grating vcsels},}\ }\href {\doibase 10.1109/JSTQE.2009.2015195} {\bibfield
  {journal} {\bibinfo  {journal} {IEEE Journal of Selected Topics in Quantum
  Electronics}\ }\textbf {\bibinfo {volume} {15}},\ \bibinfo {pages} {869--878}
  (\bibinfo {year} {2009})}\BibitemShut {NoStop}%
\bibitem [{\citenamefont {Günay-Demirkol}\ and\ \citenamefont
  {Kaya}(2012)}]{Gunay2012}%
  \BibitemOpen
  \bibfield  {author} {\bibinfo {author} {\bibfnamefont {A.}~\bibnamefont
  {Günay-Demirkol}}\ and\ \bibinfo {author} {\bibfnamefont {I.~I.}\
  \bibnamefont {Kaya}},\ }\bibfield  {title} {\enquote {\bibinfo {title}
  {Tuning of nanogap size in high tensile stress silicon nitride thin films},}\
  }\href {\doibase 10.1063/1.4712289} {\bibfield  {journal} {\bibinfo
  {journal} {Review of Scientific Instruments}\ }\textbf {\bibinfo {volume}
  {83}},\ \bibinfo {pages} {055003} (\bibinfo {year} {2012})},\ \Eprint
  {http://arxiv.org/abs/https://doi.org/10.1063/1.4712289}
  {https://doi.org/10.1063/1.4712289} \BibitemShut {NoStop}%
\bibitem [{\citenamefont {Kim}\ \emph {et~al.}(2006)\citenamefont {Kim},
  \citenamefont {Chen}, \citenamefont {Aziz}, \citenamefont {Branton},\ and\
  \citenamefont {Vlassak}}]{Kim2006}%
  \BibitemOpen
  \bibfield  {author} {\bibinfo {author} {\bibfnamefont {Y.-R.}\ \bibnamefont
  {Kim}}, \bibinfo {author} {\bibfnamefont {P.}~\bibnamefont {Chen}}, \bibinfo
  {author} {\bibfnamefont {M.~J.}\ \bibnamefont {Aziz}}, \bibinfo {author}
  {\bibfnamefont {D.}~\bibnamefont {Branton}}, \ and\ \bibinfo {author}
  {\bibfnamefont {J.~J.}\ \bibnamefont {Vlassak}},\ }\bibfield  {title}
  {\enquote {\bibinfo {title} {Focused ion beam induced deflections of
  freestanding thin films},}\ }\href {\doibase 10.1063/1.2363900} {\bibfield
  {journal} {\bibinfo  {journal} {Journal of Applied Physics}\ }\textbf
  {\bibinfo {volume} {100}},\ \bibinfo {pages} {104322} (\bibinfo {year}
  {2006})},\ \Eprint {http://arxiv.org/abs/https://doi.org/10.1063/1.2363900}
  {https://doi.org/10.1063/1.2363900} \BibitemShut {NoStop}%
\bibitem [{\citenamefont {Ajovalasit}, \citenamefont {Petrucci},\ and\
  \citenamefont {Scafidi}(2015)}]{Ajovalasit2015}%
  \BibitemOpen
  \bibfield  {author} {\bibinfo {author} {\bibfnamefont {A.}~\bibnamefont
  {Ajovalasit}}, \bibinfo {author} {\bibfnamefont {G.}~\bibnamefont
  {Petrucci}}, \ and\ \bibinfo {author} {\bibfnamefont {M.}~\bibnamefont
  {Scafidi}},\ }\bibfield  {title} {\enquote {\bibinfo {title} {Photoelastic
  analysis of edge residual stresses in glass by the automated tint plate
  method},}\ }\href {\doibase 10.1111/ext.12017} {\bibfield  {journal}
  {\bibinfo  {journal} {Experimental Techniques}\ }\textbf {\bibinfo {volume}
  {39}},\ \bibinfo {pages} {11--18} (\bibinfo {year} {2015})},\ \Eprint
  {http://arxiv.org/abs/https://onlinelibrary.wiley.com/doi/pdf/10.1111/ext.12017}
  {https://onlinelibrary.wiley.com/doi/pdf/10.1111/ext.12017} \BibitemShut
  {NoStop}%
\bibitem [{\citenamefont {Capelle}\ \emph {et~al.}(2017)\citenamefont
  {Capelle}, \citenamefont {Tsaturyan}, \citenamefont {Barg},\ and\
  \citenamefont {Schliesser}}]{Capelle2017}%
  \BibitemOpen
  \bibfield  {author} {\bibinfo {author} {\bibfnamefont {T.}~\bibnamefont
  {Capelle}}, \bibinfo {author} {\bibfnamefont {Y.}~\bibnamefont {Tsaturyan}},
  \bibinfo {author} {\bibfnamefont {A.}~\bibnamefont {Barg}}, \ and\ \bibinfo
  {author} {\bibfnamefont {A.}~\bibnamefont {Schliesser}},\ }\bibfield  {title}
  {\enquote {\bibinfo {title} {Polarimetric analysis of stress anisotropy in
  nanomechanical silicon nitride resonators},}\ }\href {\doibase
  10.1063/1.4982876} {\bibfield  {journal} {\bibinfo  {journal} {Applied
  Physics Letters}\ }\textbf {\bibinfo {volume} {110}},\ \bibinfo {pages}
  {181106} (\bibinfo {year} {2017})},\ \Eprint
  {http://arxiv.org/abs/https://doi.org/10.1063/1.4982876}
  {https://doi.org/10.1063/1.4982876} \BibitemShut {NoStop}%
\bibitem [{\citenamefont {Kemiktarak}\ \emph
  {et~al.}(2012{\natexlab{a}})\citenamefont {Kemiktarak}, \citenamefont
  {Metcalfe}, \citenamefont {Durand},\ and\ \citenamefont
  {Lawall}}]{Kemiktarak2012apl}%
  \BibitemOpen
  \bibfield  {author} {\bibinfo {author} {\bibfnamefont {U.}~\bibnamefont
  {Kemiktarak}}, \bibinfo {author} {\bibfnamefont {M.}~\bibnamefont
  {Metcalfe}}, \bibinfo {author} {\bibfnamefont {M.}~\bibnamefont {Durand}}, \
  and\ \bibinfo {author} {\bibfnamefont {J.}~\bibnamefont {Lawall}},\
  }\bibfield  {title} {\enquote {\bibinfo {title} {Mechanically compliant
  grating reflectors for optomechanics},}\ }\href {\doibase 10.1063/1.3684248}
  {\bibfield  {journal} {\bibinfo  {journal} {Applied Physics Letters}\
  }\textbf {\bibinfo {volume} {100}},\ \bibinfo {pages} {061124} (\bibinfo
  {year} {2012}{\natexlab{a}})},\ \Eprint
  {http://arxiv.org/abs/https://doi.org/10.1063/1.3684248}
  {https://doi.org/10.1063/1.3684248} \BibitemShut {NoStop}%
\bibitem [{\citenamefont {Bui}\ \emph {et~al.}(2012)\citenamefont {Bui},
  \citenamefont {Zheng}, \citenamefont {Hoch}, \citenamefont {Lee},
  \citenamefont {Harris},\ and\ \citenamefont {Wong}}]{Bui2012}%
  \BibitemOpen
  \bibfield  {author} {\bibinfo {author} {\bibfnamefont {C.~H.}\ \bibnamefont
  {Bui}}, \bibinfo {author} {\bibfnamefont {J.}~\bibnamefont {Zheng}}, \bibinfo
  {author} {\bibfnamefont {S.~W.}\ \bibnamefont {Hoch}}, \bibinfo {author}
  {\bibfnamefont {L.~Y.~T.}\ \bibnamefont {Lee}}, \bibinfo {author}
  {\bibfnamefont {J.~G.~E.}\ \bibnamefont {Harris}}, \ and\ \bibinfo {author}
  {\bibfnamefont {C.~W.}\ \bibnamefont {Wong}},\ }\bibfield  {title} {\enquote
  {\bibinfo {title} {High-reflectivity, high-q micromechanical membranes via
  guided resonances for enhanced optomechanical coupling},}\ }\href {\doibase
  10.1063/1.3658731} {\bibfield  {journal} {\bibinfo  {journal} {Applied
  Physics Letters}\ }\textbf {\bibinfo {volume} {100}},\ \bibinfo {pages}
  {021110} (\bibinfo {year} {2012})}\BibitemShut {NoStop}%
\bibitem [{\citenamefont {Kemiktarak}\ \emph
  {et~al.}(2012{\natexlab{b}})\citenamefont {Kemiktarak}, \citenamefont
  {Durand}, \citenamefont {Metcalfe},\ and\ \citenamefont
  {Lawall}}]{Kemiktarak2012njp}%
  \BibitemOpen
  \bibfield  {author} {\bibinfo {author} {\bibfnamefont {U.}~\bibnamefont
  {Kemiktarak}}, \bibinfo {author} {\bibfnamefont {M.}~\bibnamefont {Durand}},
  \bibinfo {author} {\bibfnamefont {M.}~\bibnamefont {Metcalfe}}, \ and\
  \bibinfo {author} {\bibfnamefont {J.}~\bibnamefont {Lawall}},\ }\bibfield
  {title} {\enquote {\bibinfo {title} {Cavity optomechanics with sub-wavelength
  grating mirrors},}\ }\href {http://stacks.iop.org/1367-2630/14/i=12/a=125010}
  {\bibfield  {journal} {\bibinfo  {journal} {New J. Phys.}\ }\textbf {\bibinfo
  {volume} {14}},\ \bibinfo {pages} {125010} (\bibinfo {year}
  {2012}{\natexlab{b}})}\BibitemShut {NoStop}%
\bibitem [{\citenamefont {Kemiktarak}\ \emph {et~al.}(2014)\citenamefont
  {Kemiktarak}, \citenamefont {Durand}, \citenamefont {Metcalfe},\ and\
  \citenamefont {Lawall}}]{Kemiktarak2014}%
  \BibitemOpen
  \bibfield  {author} {\bibinfo {author} {\bibfnamefont {U.}~\bibnamefont
  {Kemiktarak}}, \bibinfo {author} {\bibfnamefont {M.}~\bibnamefont {Durand}},
  \bibinfo {author} {\bibfnamefont {M.}~\bibnamefont {Metcalfe}}, \ and\
  \bibinfo {author} {\bibfnamefont {J.}~\bibnamefont {Lawall}},\ }\bibfield
  {title} {\enquote {\bibinfo {title} {Mode competition and anomalous cooling
  in a multimode phonon laser},}\ }\href {\doibase
  10.1103/PhysRevLett.113.030802} {\bibfield  {journal} {\bibinfo  {journal}
  {Phys. Rev. Lett.}\ }\textbf {\bibinfo {volume} {113}},\ \bibinfo {pages}
  {030802} (\bibinfo {year} {2014})}\BibitemShut {NoStop}%
\bibitem [{\citenamefont {Stambaugh}\ \emph {et~al.}(2015)\citenamefont
  {Stambaugh}, \citenamefont {Xu}, \citenamefont {Kemiktarak}, \citenamefont
  {Taylor},\ and\ \citenamefont {Lawall}}]{Stambaugh2015}%
  \BibitemOpen
  \bibfield  {author} {\bibinfo {author} {\bibfnamefont {C.}~\bibnamefont
  {Stambaugh}}, \bibinfo {author} {\bibfnamefont {H.}~\bibnamefont {Xu}},
  \bibinfo {author} {\bibfnamefont {U.}~\bibnamefont {Kemiktarak}}, \bibinfo
  {author} {\bibfnamefont {J.}~\bibnamefont {Taylor}}, \ and\ \bibinfo {author}
  {\bibfnamefont {J.}~\bibnamefont {Lawall}},\ }\bibfield  {title} {\enquote
  {\bibinfo {title} {From membrane-in-the-middle to mirror-in-the-middle with a
  high-reflectivity sub-wavelength grating},}\ }\href {\doibase
  10.1002/andp.201400142} {\bibfield  {journal} {\bibinfo  {journal} {Annalen
  der Physik}\ }\textbf {\bibinfo {volume} {527}},\ \bibinfo {pages} {81--88}
  (\bibinfo {year} {2015})}\BibitemShut {NoStop}%
\bibitem [{\citenamefont {Yang}\ \emph {et~al.}(2015)\citenamefont {Yang},
  \citenamefont {Gerke}, \citenamefont {Ng}, \citenamefont {Rao}, \citenamefont
  {Chase},\ and\ \citenamefont {Chang-Hasnain}}]{Yang2015}%
  \BibitemOpen
  \bibfield  {author} {\bibinfo {author} {\bibfnamefont {W.}~\bibnamefont
  {Yang}}, \bibinfo {author} {\bibfnamefont {S.~A.}\ \bibnamefont {Gerke}},
  \bibinfo {author} {\bibfnamefont {K.~W.}\ \bibnamefont {Ng}}, \bibinfo
  {author} {\bibfnamefont {Y.}~\bibnamefont {Rao}}, \bibinfo {author}
  {\bibfnamefont {C.}~\bibnamefont {Chase}}, \ and\ \bibinfo {author}
  {\bibfnamefont {C.~J.}\ \bibnamefont {Chang-Hasnain}},\ }\bibfield  {title}
  {\enquote {\bibinfo {title} {Laser optomechanics},}\ }\href
  {http://dx.doi.org/10.1038/srep13700} {\bibfield  {journal} {\bibinfo
  {journal} {Scientific Reports}\ }\textbf {\bibinfo {volume} {5}},\ \bibinfo
  {pages} {13700} (\bibinfo {year} {2015})}\BibitemShut {NoStop}%
\bibitem [{\citenamefont {Norte}, \citenamefont {Moura},\ and\ \citenamefont
  {Gr\"oblacher}(2016)}]{Norte2016}%
  \BibitemOpen
  \bibfield  {author} {\bibinfo {author} {\bibfnamefont {R.~A.}\ \bibnamefont
  {Norte}}, \bibinfo {author} {\bibfnamefont {J.~P.}\ \bibnamefont {Moura}}, \
  and\ \bibinfo {author} {\bibfnamefont {S.}~\bibnamefont {Gr\"oblacher}},\
  }\bibfield  {title} {\enquote {\bibinfo {title} {Mechanical resonators for
  quantum optomechanics experiments at room temperature},}\ }\href {\doibase
  10.1103/PhysRevLett.116.147202} {\bibfield  {journal} {\bibinfo  {journal}
  {Physical Review Letters}\ }\textbf {\bibinfo {volume} {116}},\ \bibinfo
  {pages} {147202} (\bibinfo {year} {2016})}\BibitemShut {NoStop}%
\bibitem [{\citenamefont {Bernard}\ \emph {et~al.}(2016)\citenamefont
  {Bernard}, \citenamefont {Reinhardt}, \citenamefont {Dumont}, \citenamefont
  {Peter},\ and\ \citenamefont {Sankey}}]{Bernard2016}%
  \BibitemOpen
  \bibfield  {author} {\bibinfo {author} {\bibfnamefont {S.}~\bibnamefont
  {Bernard}}, \bibinfo {author} {\bibfnamefont {C.}~\bibnamefont {Reinhardt}},
  \bibinfo {author} {\bibfnamefont {V.}~\bibnamefont {Dumont}}, \bibinfo
  {author} {\bibfnamefont {Y.-A.}\ \bibnamefont {Peter}}, \ and\ \bibinfo
  {author} {\bibfnamefont {J.~C.}\ \bibnamefont {Sankey}},\ }\bibfield  {title}
  {\enquote {\bibinfo {title} {Precision resonance tuning and design of sin
  photonic crystal reflectors},}\ }\href {\doibase 10.1364/OL.41.005624}
  {\bibfield  {journal} {\bibinfo  {journal} {Opt. Lett.}\ }\textbf {\bibinfo
  {volume} {41}},\ \bibinfo {pages} {5624--5627} (\bibinfo {year}
  {2016})}\BibitemShut {NoStop}%
\bibitem [{\citenamefont {Chen}\ \emph {et~al.}(2017)\citenamefont {Chen},
  \citenamefont {Chardin}, \citenamefont {Makles}, \citenamefont {Ca{\"e}r},
  \citenamefont {Chua}, \citenamefont {Braive}, \citenamefont {Robert-Philip},
  \citenamefont {Briant}, \citenamefont {Cohadon}, \citenamefont {Heidmann},
  \citenamefont {Jacqmin},\ and\ \citenamefont {Del{\'e}glise}}]{Chen2017}%
  \BibitemOpen
  \bibfield  {author} {\bibinfo {author} {\bibfnamefont {X.}~\bibnamefont
  {Chen}}, \bibinfo {author} {\bibfnamefont {C.}~\bibnamefont {Chardin}},
  \bibinfo {author} {\bibfnamefont {K.}~\bibnamefont {Makles}}, \bibinfo
  {author} {\bibfnamefont {C.}~\bibnamefont {Ca{\"e}r}}, \bibinfo {author}
  {\bibfnamefont {S.}~\bibnamefont {Chua}}, \bibinfo {author} {\bibfnamefont
  {R.}~\bibnamefont {Braive}}, \bibinfo {author} {\bibfnamefont
  {I.}~\bibnamefont {Robert-Philip}}, \bibinfo {author} {\bibfnamefont
  {T.}~\bibnamefont {Briant}}, \bibinfo {author} {\bibfnamefont {P.-F.}\
  \bibnamefont {Cohadon}}, \bibinfo {author} {\bibfnamefont {A.}~\bibnamefont
  {Heidmann}}, \bibinfo {author} {\bibfnamefont {T.}~\bibnamefont {Jacqmin}}, \
  and\ \bibinfo {author} {\bibfnamefont {S.}~\bibnamefont {Del{\'e}glise}},\
  }\bibfield  {title} {\enquote {\bibinfo {title} {High-finesse fabry--perot
  cavities with bidimensional si3n4 photonic-crystal slabs},}\ }\href {\doibase
  10.1038/lsa.2016.190} {\bibfield  {journal} {\bibinfo  {journal} {Light:
  Science \& Applications}\ }\textbf {\bibinfo {volume} {6}},\ \bibinfo {pages}
  {e16190} (\bibinfo {year} {2017})}\BibitemShut {NoStop}%
\bibitem [{\citenamefont {{a}o P.~Moura}\ \emph {et~al.}(2018)\citenamefont
  {{a}o P.~Moura}, \citenamefont {Norte}, \citenamefont {Guo}, \citenamefont
  {Sch\"{a}fermeier},\ and\ \citenamefont {Gr\"{o}blacher}}]{Moura2018}%
  \BibitemOpen
  \bibfield  {author} {\bibinfo {author} {\bibfnamefont {J.}~\bibnamefont {{a}o
  P.~Moura}}, \bibinfo {author} {\bibfnamefont {R.~A.}\ \bibnamefont {Norte}},
  \bibinfo {author} {\bibfnamefont {J.}~\bibnamefont {Guo}}, \bibinfo {author}
  {\bibfnamefont {C.}~\bibnamefont {Sch\"{a}fermeier}}, \ and\ \bibinfo
  {author} {\bibfnamefont {S.}~\bibnamefont {Gr\"{o}blacher}},\ }\bibfield
  {title} {\enquote {\bibinfo {title} {Centimeter-scale suspended photonic
  crystal mirrors},}\ }\href {\doibase 10.1364/OE.26.001895} {\bibfield
  {journal} {\bibinfo  {journal} {Opt. Express}\ }\textbf {\bibinfo {volume}
  {26}},\ \bibinfo {pages} {1895--1909} (\bibinfo {year} {2018})}\BibitemShut
  {NoStop}%
\bibitem [{\citenamefont {Kini~Manjeshwar}\ \emph {et~al.}(2020)\citenamefont
  {Kini~Manjeshwar}, \citenamefont {Elkhouly}, \citenamefont {Fitzgerald},
  \citenamefont {Ekman}, \citenamefont {Zhang}, \citenamefont {Zhang},
  \citenamefont {Wang}, \citenamefont {Tassin},\ and\ \citenamefont
  {Wieczorek}}]{Manjeshwar2020}%
  \BibitemOpen
  \bibfield  {author} {\bibinfo {author} {\bibfnamefont {S.}~\bibnamefont
  {Kini~Manjeshwar}}, \bibinfo {author} {\bibfnamefont {K.}~\bibnamefont
  {Elkhouly}}, \bibinfo {author} {\bibfnamefont {J.~M.}\ \bibnamefont
  {Fitzgerald}}, \bibinfo {author} {\bibfnamefont {M.}~\bibnamefont {Ekman}},
  \bibinfo {author} {\bibfnamefont {Y.}~\bibnamefont {Zhang}}, \bibinfo
  {author} {\bibfnamefont {F.}~\bibnamefont {Zhang}}, \bibinfo {author}
  {\bibfnamefont {S.~M.}\ \bibnamefont {Wang}}, \bibinfo {author}
  {\bibfnamefont {P.}~\bibnamefont {Tassin}}, \ and\ \bibinfo {author}
  {\bibfnamefont {W.}~\bibnamefont {Wieczorek}},\ }\bibfield  {title} {\enquote
  {\bibinfo {title} {Suspended photonic crystal membranes in algaas
  heterostructures for integrated multi-element optomechanics},}\ }\href
  {\doibase 10.1063/5.0012667} {\bibfield  {journal} {\bibinfo  {journal}
  {Applied Physics Letters}\ }\textbf {\bibinfo {volume} {116}},\ \bibinfo
  {pages} {264001} (\bibinfo {year} {2020})},\ \Eprint
  {http://arxiv.org/abs/https://doi.org/10.1063/5.0012667}
  {https://doi.org/10.1063/5.0012667} \BibitemShut {NoStop}%
\bibitem [{\citenamefont {Guo}, \citenamefont {Norte},\ and\ \citenamefont
  {Gr\"{o}blacher}(2017)}]{Guo2017}%
  \BibitemOpen
  \bibfield  {author} {\bibinfo {author} {\bibfnamefont {J.}~\bibnamefont
  {Guo}}, \bibinfo {author} {\bibfnamefont {R.~A.}\ \bibnamefont {Norte}}, \
  and\ \bibinfo {author} {\bibfnamefont {S.}~\bibnamefont {Gr\"{o}blacher}},\
  }\bibfield  {title} {\enquote {\bibinfo {title} {Integrated optical force
  sensors using focusing photonic crystal arrays},}\ }\href {\doibase
  10.1364/OE.25.009196} {\bibfield  {journal} {\bibinfo  {journal} {Opt.
  Express}\ }\textbf {\bibinfo {volume} {25}},\ \bibinfo {pages} {9196--9203}
  (\bibinfo {year} {2017})}\BibitemShut {NoStop}%
\bibitem [{\citenamefont {Naesby}\ and\ \citenamefont
  {Dantan}(2018)}]{Naesby2018}%
  \BibitemOpen
  \bibfield  {author} {\bibinfo {author} {\bibfnamefont {A.}~\bibnamefont
  {Naesby}}\ and\ \bibinfo {author} {\bibfnamefont {A.}~\bibnamefont
  {Dantan}},\ }\bibfield  {title} {\enquote {\bibinfo {title} {Microcavities
  with suspended subwavelength structured mirrors},}\ }\href {\doibase
  10.1364/OE.26.029886} {\bibfield  {journal} {\bibinfo  {journal} {Opt.
  Express}\ }\textbf {\bibinfo {volume} {26}},\ \bibinfo {pages} {29886--29894}
  (\bibinfo {year} {2018})}\BibitemShut {NoStop}%
\bibitem [{\citenamefont {G\"{a}rtner}\ \emph {et~al.}(2018)\citenamefont
  {G\"{a}rtner}, \citenamefont {Moura}, \citenamefont {Haaxman}, \citenamefont
  {Norte},\ and\ \citenamefont {Gr\"{o}blacher}}]{Gartner2018}%
  \BibitemOpen
  \bibfield  {author} {\bibinfo {author} {\bibfnamefont {C.}~\bibnamefont
  {G\"{a}rtner}}, \bibinfo {author} {\bibfnamefont {J.~P.}\ \bibnamefont
  {Moura}}, \bibinfo {author} {\bibfnamefont {W.}~\bibnamefont {Haaxman}},
  \bibinfo {author} {\bibfnamefont {R.~A.}\ \bibnamefont {Norte}}, \ and\
  \bibinfo {author} {\bibfnamefont {S.}~\bibnamefont {Gr\"{o}blacher}},\
  }\bibfield  {title} {\enquote {\bibinfo {title} {Integrated optomechanical
  arrays of two high reflectivity sin membranes},}\ }\href
  {https://www.ncbi.nlm.nih.gov/pubmed/30247926} {\bibfield  {journal}
  {\bibinfo  {journal} {Nano letters}\ }\textbf {\bibinfo {volume} {18}},\
  \bibinfo {pages} {7171--7175} (\bibinfo {year} {2018})}\BibitemShut {NoStop}%
\bibitem [{\citenamefont {\ifmmode~\check{C}\else \v{C}\fi{}ernot\'{\i}k},
  \citenamefont {Dantan},\ and\ \citenamefont {Genes}(2019)}]{Cernotik2019}%
  \BibitemOpen
  \bibfield  {author} {\bibinfo {author} {\bibfnamefont {O.}~\bibnamefont
  {\ifmmode~\check{C}\else \v{C}\fi{}ernot\'{\i}k}}, \bibinfo {author}
  {\bibfnamefont {A.}~\bibnamefont {Dantan}}, \ and\ \bibinfo {author}
  {\bibfnamefont {C.}~\bibnamefont {Genes}},\ }\bibfield  {title} {\enquote
  {\bibinfo {title} {Cavity quantum electrodynamics with frequency-dependent
  reflectors},}\ }\href {\doibase 10.1103/PhysRevLett.122.243601} {\bibfield
  {journal} {\bibinfo  {journal} {Phys. Rev. Lett.}\ }\textbf {\bibinfo
  {volume} {122}},\ \bibinfo {pages} {243601} (\bibinfo {year}
  {2019})}\BibitemShut {NoStop}%
\bibitem [{\citenamefont {Dantan}(2020)}]{Dantan2020}%
  \BibitemOpen
  \bibfield  {author} {\bibinfo {author} {\bibfnamefont {A.}~\bibnamefont
  {Dantan}},\ }\bibfield  {title} {\enquote {\bibinfo {title} {Membrane
  sandwich squeeze film pressure sensors},}\ }\href {\doibase
  10.1063/5.0011795} {\bibfield  {journal} {\bibinfo  {journal} {Journal of
  Applied Physics}\ }\textbf {\bibinfo {volume} {128}},\ \bibinfo {pages}
  {091101} (\bibinfo {year} {2020})},\ \Eprint
  {http://arxiv.org/abs/https://doi.org/10.1063/5.0011795}
  {https://doi.org/10.1063/5.0011795} \BibitemShut {NoStop}%
\bibitem [{\citenamefont {Hübner}\ \emph {et~al.}(2003)\citenamefont
  {Hübner}, \citenamefont {Morgenroth}, \citenamefont {Meyer}, \citenamefont
  {Sulzbach}, \citenamefont {Brendel},\ and\ \citenamefont
  {Mirandé}}]{Hubner2003}%
  \BibitemOpen
  \bibfield  {author} {\bibinfo {author} {\bibfnamefont {U.}~\bibnamefont
  {Hübner}}, \bibinfo {author} {\bibfnamefont {W.}~\bibnamefont {Morgenroth}},
  \bibinfo {author} {\bibfnamefont {H.~G.}\ \bibnamefont {Meyer}}, \bibinfo
  {author} {\bibfnamefont {T.}~\bibnamefont {Sulzbach}}, \bibinfo {author}
  {\bibfnamefont {B.}~\bibnamefont {Brendel}}, \ and\ \bibinfo {author}
  {\bibfnamefont {W.}~\bibnamefont {Mirandé}},\ }\bibfield  {title} {\enquote
  {\bibinfo {title} {Downwards to metrology in nanoscale: determination of the
  afm tip shape with well-known sharp-edged calibration structures},}\ }\href
  {\doibase 10.1007/s00339-002-1975-6} {\bibfield  {journal} {\bibinfo
  {journal} {Applied Physics A}\ }\textbf {\bibinfo {volume} {76}},\ \bibinfo
  {pages} {913--917} (\bibinfo {year} {2003})}\BibitemShut {NoStop}%
\bibitem [{\citenamefont {{Lianqing Liu}}\ \emph {et~al.}(2008)\citenamefont
  {{Lianqing Liu}}, \citenamefont {{Ning Xi}}, \citenamefont {{Jiangbo Zhang}},
  \citenamefont {{Guangyong Li}}, \citenamefont {{Yuechao Wang}},\ and\
  \citenamefont {{Zaili Dong}}}]{Lianqing2008}%
  \BibitemOpen
  \bibfield  {author} {\bibinfo {author} {\bibnamefont {{Lianqing Liu}}},
  \bibinfo {author} {\bibnamefont {{Ning Xi}}}, \bibinfo {author} {\bibnamefont
  {{Jiangbo Zhang}}}, \bibinfo {author} {\bibnamefont {{Guangyong Li}}},
  \bibinfo {author} {\bibnamefont {{Yuechao Wang}}}, \ and\ \bibinfo {author}
  {\bibnamefont {{Zaili Dong}}},\ }\bibfield  {title} {\enquote {\bibinfo
  {title} {System positioning error compensated by local scan in atomic force
  microscope based nanomanipulation},}\ }in\ \href {\doibase
  10.1109/NEMS.2008.4484513} {\emph {\bibinfo {booktitle} {2008 3rd IEEE
  International Conference on Nano/Micro Engineered and Molecular Systems}}}\
  (\bibinfo {year} {2008})\ pp.\ \bibinfo {pages} {1113--1118}\BibitemShut
  {NoStop}%
\bibitem [{\citenamefont {Germer}()}]{MIST}%
  \BibitemOpen
  \bibfield  {author} {\bibinfo {author} {\bibfnamefont {T.}~\bibnamefont
  {Germer}},\ }\href@noop {} {\emph {\bibinfo {title} {Modeled Integrated
  Scatter Tool version 3.01}}}\BibitemShut {NoStop}%
\bibitem [{\citenamefont {Fan}\ and\ \citenamefont
  {Joannopoulos}(2002)}]{Fan2002}%
  \BibitemOpen
  \bibfield  {author} {\bibinfo {author} {\bibfnamefont {S.}~\bibnamefont
  {Fan}}\ and\ \bibinfo {author} {\bibfnamefont {J.~D.}\ \bibnamefont
  {Joannopoulos}},\ }\bibfield  {title} {\enquote {\bibinfo {title} {Analysis
  of guided resonances in photonic crystal slabs},}\ }\href {\doibase
  10.1103/PhysRevB.65.235112} {\bibfield  {journal} {\bibinfo  {journal} {Phys.
  Rev. B}\ }\textbf {\bibinfo {volume} {65}},\ \bibinfo {pages} {235112}
  (\bibinfo {year} {2002})}\BibitemShut {NoStop}%
\bibitem [{\citenamefont {Bykov}\ and\ \citenamefont
  {Doskolovich}(2015)}]{Bykov2015}%
  \BibitemOpen
  \bibfield  {author} {\bibinfo {author} {\bibfnamefont {D.~A.}\ \bibnamefont
  {Bykov}}\ and\ \bibinfo {author} {\bibfnamefont {L.~L.}\ \bibnamefont
  {Doskolovich}},\ }\bibfield  {title} {\enquote {\bibinfo {title}
  {Spatiotemporal coupled-mode theory of guided-mode resonant gratings},}\
  }\href {\doibase 10.1364/OE.23.019234} {\bibfield  {journal} {\bibinfo
  {journal} {Opt. Express}\ }\textbf {\bibinfo {volume} {23}},\ \bibinfo
  {pages} {19234--19241} (\bibinfo {year} {2015})}\BibitemShut {NoStop}%
\bibitem [{\citenamefont {Timoshenko}\ and\ \citenamefont
  {Woinowsky-Krieger}(1959)}]{Timoshenko1959}%
  \BibitemOpen
  \bibfield  {author} {\bibinfo {author} {\bibfnamefont {S.}~\bibnamefont
  {Timoshenko}}\ and\ \bibinfo {author} {\bibfnamefont {S.}~\bibnamefont
  {Woinowsky-Krieger}},\ }\href@noop {} {\emph {\bibinfo {title} {Theory of
  Plates and Shells}}}\ (\bibinfo  {publisher} {MacGraw-Hill, New York},\
  \bibinfo {year} {1959})\BibitemShut {NoStop}%
\bibitem [{Zie(2014)}]{Zienkiewicz2014}%
  \BibitemOpen
  \bibfield  {title} {\enquote {\bibinfo {title} {The finite element method for
  solid and structural mechanics},}\ }in\ \href {\doibase
  https://doi.org/10.1016/B978-1-85617-634-7.00016-8} {\emph {\bibinfo
  {booktitle} {The Finite Element Method for Solid and Structural Mechanics
  (Seventh Edition)}}},\ \bibinfo {editor} {edited by\ \bibinfo {editor}
  {\bibfnamefont {O.}~\bibnamefont {Zienkiewicz}}, \bibinfo {editor}
  {\bibfnamefont {R.}~\bibnamefont {Taylor}}, \ and\ \bibinfo {editor}
  {\bibfnamefont {D.}~\bibnamefont {Fox}}}\ (\bibinfo  {publisher}
  {Butterworth-Heinemann},\ \bibinfo {address} {Oxford},\ \bibinfo {year}
  {2014})\ \bibinfo {edition} {seventh edition}\ ed.\BibitemShut {Stop}%
\end{thebibliography}%

\end{document}